\documentclass[12pt, letterpaper]{article}
\usepackage{geometry}
\usepackage{amsmath,amssymb,amsfonts}
\usepackage[onehalfspacing]{setspace}
\usepackage{abstract}
\usepackage{natbib,bibmods}
\usepackage{enumerate,enumitem}
\usepackage{url}
\usepackage{libertine}
\usepackage{adjustbox}
\usepackage[T1]{fontenc}
\usepackage{soul}
\usepackage[usenames,dvipsnames]{color}
\usepackage{xcolor,colortbl}
\usepackage{threeparttable,multirow}
\usepackage[colorlinks,urlcolor=Blue,citecolor=Blue, linkcolor=Blue]{hyperref}

\usepackage{caption}
\usepackage{subcaption}
\usepackage{ushort}
\usepackage{xfrac}
\usepackage{threeparttable}
\usepackage{lscape}
\usepackage{rotating}
\usepackage{pgf,tikz}
\usetikzlibrary{arrows,decorations.markings}
\usepackage{tabularx,ragged2e,booktabs}
\usepackage{graphicx} 
\usepackage{eurosym}
\usepackage{siunitx}
\usepackage{ntheorem}
\usepackage{float}
\usepackage{changepage}
\usepackage{bbm}

\begin{document}

\title{\begin{spacing}{1}Spousal Occupational Sorting and COVID-19 Incidence:\\Evidence from the United States\end{spacing}}

\author{Egor Malkov\footnote{~E-mail: \href{malko017@umn.edu}{malko017@umn.edu}. ORCID: 0000-0002-4566-5853. I am grateful to Milena Almagro, Adam Blandin, Jeffrey Clemens, Anna Stansbury, Philippe Van Kerm, Mariana Viollaz, and several anonymous referees for very useful feedback. This is a substantially revised part of the paper appeared in Issue 34 of CEPR's \textit{Covid Economics: Vetted and Real-Time Papers} under title ``Nature of Work and Distribution of Risk: Evidence from Occupational Sorting, Skills, and Tasks''. The views expressed herein are those of the author and not necessarily those of the Federal Reserve Bank of Minneapolis or the Federal Reserve System.}}
\date{\begin{spacing}{1} Department of Economics\\University of Minnesota\\~\\Federal Reserve Bank of Minneapolis\\~\\This version: June 2021\end{spacing}}

\maketitle

\begin{spacing}{1.0}
\begin{abstract}
\noindent How do matching of spouses and the nature of work jointly shape the distribution of COVID-19 health risks? To address this question, I study the association between the incidence of COVID-19 and the degree of spousal sorting into occupations that differ by contact intensity at the workplace. The mechanism, that I explore, implies that the higher degree of positive spousal sorting mitigates intra-household contagion and this translates into a smaller number of individuals exposed to COVID-19 risk. Using the U.S. data at the state level, I argue that spousal sorting is an important factor for understanding the disparities in the prevalence of COVID-19 during the early stages of the pandemic. First, I document that it creates about two-thirds of the U.S. dual-earner couples that are exposed to higher COVID-19 health risk due to within-household transmission. Moreover, I uncover substantial heterogeneity in the degree of spousal sorting by state. Next, for the first week of April 2020, I estimate that a one standard deviation increase in the measure of spousal sorting is associated with a 30\% reduction in the total number of cases per 100000 inhabitants and a 39.3\% decline in the total number of deaths per 100000 inhabitants. Furthermore, I find substantial temporal heterogeneity as the coefficients decline in magnitude over time. My results speak to the importance of policies that allow mitigating intra-household contagion.
\medskip

\smallskip

\end{abstract}
\end{spacing}

\newpage

\section{Introduction}\label{Introduction}

Coronavirus disease 2019 (COVID-19) pandemic created substantial challenges for health systems and economies all over the world. To reduce the spread of disease, many countries imposed numerous mitigation measures, such as lockdowns and stay-at-home orders. These policies forced many people to work from home. However, a sizeable fraction of jobs, e.g. in the United States it is equal to 21.6\% of the workforce \citep{leibovici2020social}, requires high contact intensity at the workplace and cannot be performed remotely. The nature of work became a crucial factor behind the exposure to health, income, and unemployment risks.

As for the health risk, the workers whose jobs do not require high contact intensity at the workplace face a lower risk of being infected, compared to those who work in high physical proximity to the other individuals \citep{covid19clusters, mutambudzi2021occupation}. In turn, the members of these workers' households are less exposed to the risk of within-household COVID-19 transmission in the former than in the latter case. In other words, the patterns of intra-household contagion depend on the joint distribution of spouses by occupation: both own and spousal degree of contact intensity at the job matter.

In this paper, I evaluate the importance of spousal occupational sorting in explaining the disparities in COVID-19 incidence across the United States. Following the aforementioned idea, if at least one of the spouses in a dual-earner couple works in a high contact intensity occupation, then the other family members are more exposed to the risk of being infected. Therefore, for the population to be less exposed to the intra-household contagion risk, the fraction of couples with exactly one high contact intensity worker should be lower, while the fraction of couples where both spouses have low contact intensity jobs should be higher.\footnote{~In this case, the COVID-19 health risk will be concentrated in couples where both spouses have high contact intensity jobs, but this risk has its roots in the nature of work and will not be amplified by intra-household contagion.} In other words, \textit{higher degree of positive occupational sorting} is associated with \textit{smaller number of individuals who are exposed to COVID-19 health risk}. Two regions characterized by identical distributions of males and females by occupation may demonstrate substantially different exposure to COVID-19 contagion risk depending on the patterns in spousal sorting.

To address the question about the relevance of this mechanism, I construct a state-level measure of spousal sorting using the 2015-2019 American Community Survey (ACS) individual data merged with the classification of occupations by contact intensity from \cite{mongey2021which}. In particular, I use the correlation between the contact intensity degrees of husband and wife's occupations. Next, I combine this measure with the state-level daily data on COVID-19 cases and deaths at the early stages of the pandemic, namely, April-June, 2020, provided by the Johns Hopkins University Center for Systems Science and Engineering \citep{dong2020interactive}, and demographic and socioeconomic characteristics of the states from the 2015-2019 ACS. As for the latter, I use the data on racial, age, and gender composition, the share of married couples, average household size, population density, income, employment by occupation groups, and commuting patterns.

At the aggregate level, spousal occupational sorting in the United States creates 64.2\% dual-earner couples with at least one spouse whose job requires high contact intensity at the workplace. These families are exposed to greater intra-household contagion risk. I also estimate this fraction under two counterfactual distributions: zero sorting (or random matching of spouses) and ``ideal'' sorting (the distribution with maximum feasible positive sorting). Under zero sorting, it is equal to 66.5\%, under ideal sorting---48.3\%. Furthermore, I show that the aggregate distribution masks significant state-level heterogeneity with the District of Columbia has the highest and North Dakota has the lowest degree of spousal sorting.

In my empirical analysis, I focus on the relationship between
the total number of COVID-19 cases and deaths per 100000 inhabitants and the measure of spousal occupational sorting by state. I run the regressions week by week, hence allowing the coefficients to be time-varying. In all the regressions, I control for the battery of demographic and socioeconomic variables that are widely considered as potential factors of COVID-19 spread and include day fixed effects to capture the common factors across all the states. I find that in the week of April 1-7 a one standard deviation increase in the measure of spousal sorting (this corresponds to moving from Oregon to New York) is associated with a decrease in the cumulative number of cases per 100000 inhabitants by 30\%. This represents 21.5 fewer cases per 100000 inhabitants from a sample mean of 72.4 per 100000. As for the number of deaths, I estimate that a one standard deviation increase in the measure of spousal sorting is associated with a decline in the cumulative number of deaths per 100000 inhabitants by 39.3\%. This represents 0.9 fewer cases per 100000 inhabitants from a sample mean of 2.2 per 100000. I find that these effects are stronger at the early stage of the pandemic as both coefficients decrease in magnitude over time. Furthermore, in the regressions for the number of cases, spousal sorting becomes insignificant starting from the week of April 22-28, while in the regressions for the number of deaths it is significant until the week of June 17-23. To the best of my knowledge, my paper is the first one that documents the important role of spousal occupational sorting in the incidence of COVID-19.\footnote{~\cite{lekfuangfu2020covid} and \cite{piyapromdee2020income} discuss the relationship between spousal sorting and labor market risks in the context of COVID-19 pandemic for Thailand and the United Kingdom.}

My findings about the effects of the other variables are in line with the existing research. In particular, I show that the share of males is positively correlated and the share of married couples is negatively correlated with the number of cases and deaths per capita.\footnote{~This is consistent with the results by \cite{bwire2020coronavirus}, \cite{drefahl2020population}, and \cite{peckham2020male}. \cite{bwire2020coronavirus} points out that a significant part of the gap in the number of deaths between men and women is explained by the difference in gender behavior, e.g. men tend to smoke and drink more than women, and women are more likely to take preventive measures, such as handwashing and wearing a mask.} Furthermore, I find that population density and the use of public transportation are associated with the higher number of cases and deaths per capita.\footnote{~See the discussion of these factors by \cite{almagro2020nyc} and \cite{glaeser2021much} for major U.S. cities.}

The main lesson from my findings is that the interaction between the nature of work (distribution of workers by occupations of different contact intensity) and patterns in spousal sorting plays an important role in explaining COVID-19 incidence across the United States.\footnote{~In the previous version of this paper, I also discuss the role of spousal occupational sorting in shaping income risk during the COVID-19 pandemic.} Furthermore, my results suggest several policy implications. First, targeting individuals who work in occupations that require high contact intensity with testing and vaccination, and providing them with protective equipment, would allow for reducing the health risk not only for these workers but also for the members of their households. This is an indirect way of mitigating the intra-household contagion channel. On the other hand, it is also necessary to think about the direct measures aimed at mitigating the within-household transmission. For example, providing shelter for high contact intensity workers will likely reduce the risk of contracting a disease for their families. The scope of these policies is quite sizable since, as I document, about two-thirds of the U.S. dual-earner couples are exposed to COVID-19 health risk through intra-household contagion.

My paper is related to voluminous literature studying the characteristics that account for disparities in the prevalence of COVID-19, such as local policies \citep{aparicio2021covid, berry2021evaluating}, demographic characteristics \citep{harris2020data, borjas2020demographic, levin2020assessing, sa2020socioeconomic}, commuting patterns \citep{glaeser2021much}, and occupations \citep{almagro2020nyc}. I complement these studies by showing that spousal occupational sorting is a significant factor of COVID-19 incidence even after controlling for these variables.

More broadly, my results confirm the idea that many effects of the pandemic are mediated through the economics of the household \citep{davis2021many}. In particular, this paper bridges the medical literature on the important role of within-household transmission of COVID-19 \citep{grijalva2020transmission, lei2020household, li2020characteristics, madewell2020household} to the literature on family economics in the context of the current pandemic \citep{albanesi2021gendered, alon2020impact, heggeness2020estimating, peluffo2021intra, tertilt2020time}. Furthermore, my paper is related to the literature that studies the primary role of the nature of work---teleworkability and contact intensity---in the current pandemic \citep{dingel2020many, leibovici2020social, mongey2021which}.

The rest of the paper is organized as follows. In Section \ref{Data}, I describe the data and construction of the variables. In Section \ref{Empirical Results}, I discuss the sorting of spouses by occupation contact intensity in the United States and provide the results describing its effects on COVID-19 incidence. Section \ref{Conclusion} concludes.

\section{Data}\label{Data}

I study the period between April 1, 2020 and July 1, 2020. Most of the U.S. states imposed restrictions, such as stay-at-home orders, school closures, and suspension of public gatherings, in late March 2020.\footnote{~Nine remaining states---Florida, Georgia, Maine, Mississippi, Missouri, Nevada, Pennsylvania, South Carolina, and Texas---imposed the restrictions during the first week of April 2020. Source: \href{https://github.com/govex/COVID-19/blob/master/data_tables/restrictions_policies_detailed.csv}{https://github.com/govex/COVID-19/blob/master/data\_tables/restrictions\_policies\_detailed.csv} (accessed on June 19, 2021).} My analysis starts at the point when the measures against COVID-19 contagion in the public places already went into effect, hence the results are unlikely to be affected by differential timing of these policies. Furthermore, I limit my time frame to April, May, and June 2020 because I want to guarantee that the distribution of workers by occupations and hence the measure of spousal occupational sorting are as precise as possible. Given that the U.S. labor market collapsed across all the states, and almost all occupations and industries irrespective of the state-level policies imposed \citep{kahn2020demand}, it is natural to expect that in the second half of 2020 the actual occupation shares substantially differ from the pre-pandemic shares that I use. At the same time, I do not expect that demographic characteristics of the states significantly change over the course of the pandemic.

To conduct the empirical analysis, I combine several datasets. First, I use the data on pre-pandemic demographic and socioeconomic characteristics of the states from the 2015-2019 American Community Survey. For comparability purposes, I include the characteristics considered in the methodologically close papers that study the factors affecting the prevalence of COVID-19 \citep{almagro2020nyc, borjas2020demographic}. In particular, I use the five-year estimates of the state-level average household size, share of married couples, share of males, shares of different age groups (20-39, 40-59, and above 60), shares of Black, Hispanic, and Asian population, share of people that do not have health insurance, median income, population density, average commute time to work, and the share of people who use public transportation. Finally, motivated by the findings from \cite{almagro2020nyc} about the important role of occupations in explaining the disparities in COVID-19 incidence, I also control for the employment shares of working-age population by occupation. In particular, following their classification, I construct thirteen occupation groups that capture different characteristics of the nature of work:
\begin{enumerate}
\item \textit{Essential - Professional:} Management, business, and financial occupations (OCC codes 0010-0960 in the 2018-onward Census occupational classification system).
\item \textit{Non-Essential - Professional:} Computer and mathematical occupations (OCC 1005-1240), Architecture and engineering occupations (OCC 1305-1560), Community and social service occupations (OCC 2001-2060), Educational instruction, and library occupations (OCC 2205-2555), Arts, design, entertainment, sports, and media (OCC 2600-2920), Office and administrative support occupations (OCC 5000-5940), Sales and related (OCC 4700-4965).
\item \textit{Science:} Life, physical, and social science occupations (OCC 1600-1980).
\item \textit{Law and Related:} Legal occupations (OCC 2100-2180).
\item \textit{Healthcare Practitioners:} Health diagnosing and treating practitioners and other technical occupations (OCC 3000-3270).
\item \textit{Other Health:} Health technologists and technicians (OCC 3300-3550).
\item \textit{Firefighting:} Firefighting and prevention, and other protective service workers including supervisors (OCC 3700-3750).
\item \textit{Law Enforcement:} Law enforcement workers including supervisors (OCC 3801-3960).
\item \textit{Essential - Service:} Food preparation and serving related occupations (OCC 4000-4160), Building and grounds cleaning and maintenance occupations (OCC 4200-4255).
\item \textit{Non-Essential - Service:} Personal care and service occupations (OCC 4330-4655).
\item \textit{Industrial and Construction:} Farming, fishing, and forestry (OCC 6005-6130), Construction and extraction (OCC 6200-6950), Material moving (OCC 9510-9760).
\item \textit{Essential - Technical:} Installation, maintenance, and repair occupations (OCC 7000-7640).
\item \textit{Transportation:} Transportation occupations (OCC 9005-9430).
\end{enumerate}

Second, I use the 2015-2019 ACS individual-level data to construct the measure of spousal occupational sorting.\footnote{~The data is extracted from IPUMS at \href{https://usa.ipums.org/usa/}{https://usa.ipums.org/usa/}.} The sample is constrained to dual-earner married couples where both spouses are aged 20 to 65. I use the information on occupation of each individual, and determine whether it requires high contact intensity at the workplace or not. I take the classification of occupations by contact intensity from \cite{mongey2021which}. Using ``Physical Proximity'' from O*NET Work Context module as an input, they construct a binary measure of contact intensity at the 3-digit Census OCC level. I complement their classification by manually adding occupations that they do not characterize. I define an occupation to be \textit{low CI} (``low contact intensity'') if it is classified by \cite{mongey2021which} as not requiring high contact intensity and \textit{high CI} (``high contact intensity'') if it requires high contact intensity at the workplace.

\begin{spacing}{1}
	\begin{table}[t!]
		\begin{center}
			\caption{Summary statistics} \label{tab: summary_statistics}
				\begin{tabular}{lccccc}
					\hline \hline
					Variable & Mean & St. Dev. & P10 & Median & P90 \\
					\hline
					Spousal Occupational Sorting & 0.081 & 0.022 & 0.060 & 0.080 & 0.107 \\
					Household Size & 2.56 & 0.17 & 2.39 & 2.52 & 2.70 \\
					Married Couples (\%) & 48.2 & 4.3 & 44.4 & 48.3 & 51.2 \\
					Males (\%) & 49.4 & 0.8 & 48.5 & 49.3 & 50.3 \\
					Age 20-39 (\%) & 27.0 & 2.3 & 25.2 & 26.6 & 29.0 \\
					Age 40-59 (\%) & 25.5 & 1.4 & 23.9 & 25.7 & 26.8 \\
					Age 60 and Above (\%) & 22.4 & 2.4 & 19.7 & 22.4 & 25.0 \\
					Black (\%) & 11.3 & 10.7 & 1.4 & 7.6 & 26.8 \\
					Hispanic (\%) & 11.9 & 10.3 & 3.7 & 9.4 & 25.6 \\
					Asian (\%) & 4.2 & 5.5 & 1.4 & 2.8 & 8.2 \\
					No health insurance (\%) & 8.1 & 3.1 & 4.5 & 7.9 & 12.3 \\
					Median Income (2019 USD) & 31622 & 4513 & 26295 & 31033 & 36076 \\
					Population Density (per sq. mile) & 419 & 1583 & 17.3 & 108 & 620 \\
					Avg. Commute Time (minutes) & 24.8 & 4.0 & 19.1 & 24.8 & 29.8 \\
					Public Transport Usage (\%) & 3.7 & 6.3 & 0.5 & 1.3 & 8.4 \\
					Essential - Professional (\%) & 15.4 & 2.4 & 13.1 & 15.2 & 18.0 \\
					Non-Essential - Professional (\%) & 36.1 & 1.7 & 33.7 & 36.0 & 38.2 \\
					Science (\%) & 1.0 & 0.5 & 0.6 & 0.9 & 1.5 \\
					Law and Related (\%) & 1.1 & 0.9 & 0.7 & 0.9 & 1.3 \\
					Healthcare Practitioners (\%) & 6.2 & 0.8 & 5.4 & 6.3 & 6.9 \\
					Other Health (\%) & 5.2 & 0.8 & 4.2 & 5.1 & 6.0 \\
					Firefighting (\%) & 1.1 & 0.3 & 0.8 & 1.1 & 1.5 \\
					Law Enforcement (\%) & 0.9 & 0.2 & 0.7 & 0.9 & 1.2 \\
					Essential - Service (\%) & 9.6 & 1.3 & 8.6 & 9.4 & 10.7 \\
					Non-Essential - Service (\%) & 2.8 & 0.4 & 2.5 & 2.7 & 3.0 \\
					Industrial and Construction (\%) & 9.7 & 1.6 & 7.7 & 9.9 & 11.2 \\
					Essential - Technical (\%) & 3.2 & 0.6 & 2.6 & 3.3 & 3.9 \\
					Transportation (\%) & 3.8 & 0.5 & 3.3 & 3.8 & 4.4 \\
					\hline \hline
				\end{tabular}
			\justify\footnotesize{\textsc{Notes}: The state-level data is from the ACS five-year estimates from 2015-2019. To construct the measure of spousal occupational sorting, I use the 2015-2019 ACS individual data matched with contact intensity classification from \cite{mongey2021which}. The groups of occupations are from \cite{almagro2020nyc}.}
		\end{center}
	\end{table}
\end{spacing}
\medskip

Next, I describe the construction of spousal occupational sorting measure. From the ACS individual data merged with the classification from \cite{mongey2021which}, I obtain the distribution of spouses in dual-earner couples by contact intensity of occupations. Denote the fraction of couples in state $s$ where both spouses have low CI (correspondingly, high CI) jobs by $\pi^s_{ll}$ (correspondingly, $\pi^s_{hh}$). Next, denote the fraction of couples where a male has a high CI job and a female has a low CI job (correspondingly, a male has a low CI job and a female has a high CI job) by $\pi^s_{hl}$ (correspondingly, $\pi^s_{lh}$). By definition, $\pi^s_{hh} + \pi^s_{ll} + \pi^s_{hl} + \pi^s_{lh} = 1$ for each state $s$. Next, denote the share of females (correspondingly, males) in state $s$ employed in the high CI occupations by $q^{f,s}_h$ (correspondingly, $q^{m,s}_h$). To measure the level of spousal occupational sorting in state $s$, I use cross-sectional Pearson correlation between the contact intensity degrees of husband and wife's occupations. This correlation is given by
\begin{equation}
\rho_s = \frac{\pi^s_{hh} \pi^s_{ll} - \pi^s_{hl} \pi^s_{lh}}{\sqrt{q^{m,s}_h \left( 1 - q^{m,s}_h \right) q^{f,s}_h \left( 1 - q^{f,s}_h \right)}}
\label{eq: occ_sorting_correlation}
\end{equation}

Finally, I merge the demographic and socioeconomic variables obtained from the ACS with the data on COVID-19 incidence. In particular, I use the daily state-level data on COVID-19 cases and deaths from the Johns Hopkins University Center for Systems Science and Engineering \citep{dong2020interactive}. In Table \ref{tab: summary_statistics}, I report the summary statistics for state characteristics. Notably, spousal correlation has a mean value of 0.081, and varies from 0.004 (North Dakota) to 0.141 (District of Columbia). In Table \ref{tab: spousal_sorting_states}, I report the distribution of couples by contact intensity of occupations and resulting measure of spousal sorting by state.

\section{Empirical Results}\label{Empirical Results}

In this section, I begin by describing the mechanism through which the distribution of couples by contact intensity of occupations may affect the prevalence of COVID-19. Second, I discuss the patterns of spousal occupational sorting in the United States at the aggregate and state levels. Finally, using the state-level variation, I show that spousal occupational sorting is an important factor for understanding the disparities in COVID-19 incidence even after controlling for the rich set of other characteristics that are commonly considered as drivers of COVID-19 contagion risk.

\subsection{Mechanism}\label{Mechanism}

\begin{figure}[t!]
	\centering
	\begin{subfigure}{.90\textwidth}
		\centering
		\includegraphics[width=\linewidth]{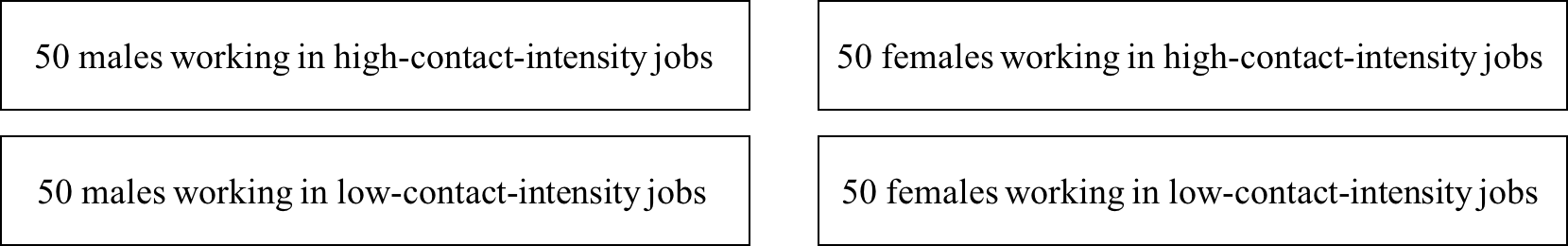}
		\subcaption{Distribution of males and females by occupations of different contact intensity}
		\label{fig: figure_1_ci}
	\end{subfigure}
	\bigskip
	\bigskip
	
	\begin{subfigure}{.32\textwidth}
		\centering
		\includegraphics[width=\linewidth]{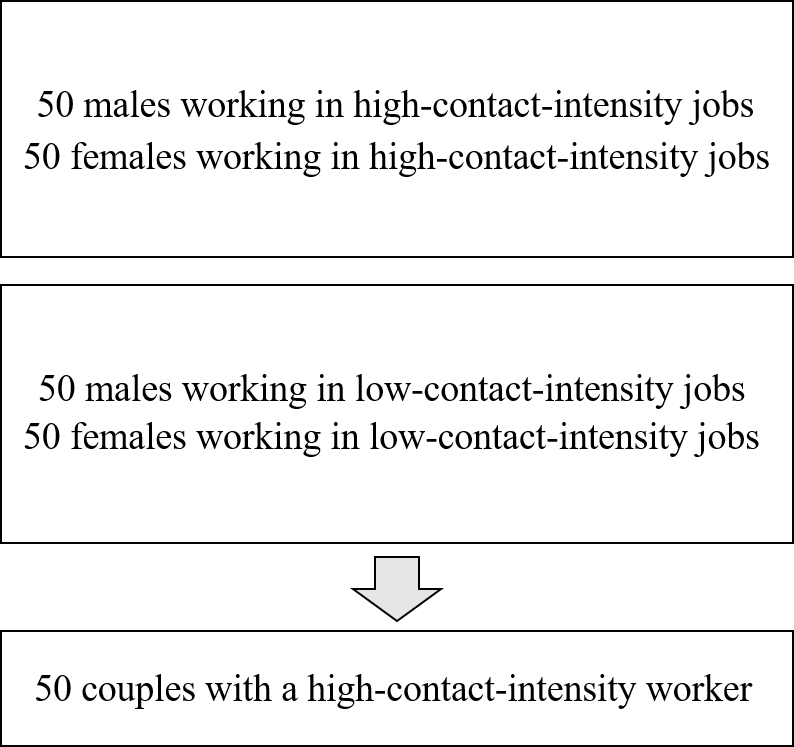}
		\subcaption{Perfect positive sorting}
		\label{fig: figure_3_ci}
	\end{subfigure}%
	$~~$
	\begin{subfigure}{.32\textwidth}
		\centering
		\includegraphics[width=\linewidth]{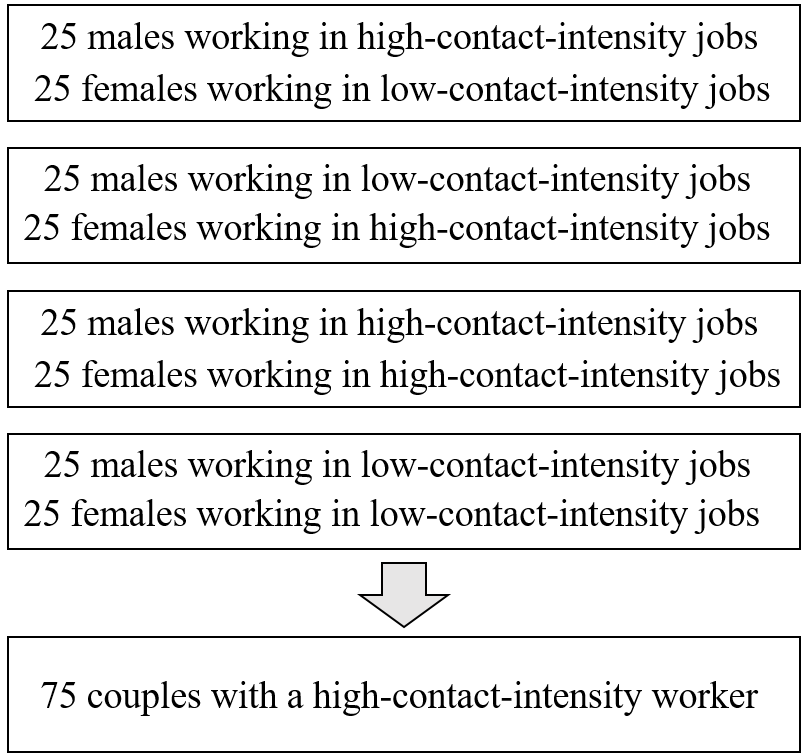}
		\subcaption{Zero sorting}
		\label{fig: figure_2-5_ci}
	\end{subfigure}%
	$~~$
	\begin{subfigure}{.32\textwidth}
		\centering
		\includegraphics[width=\linewidth]{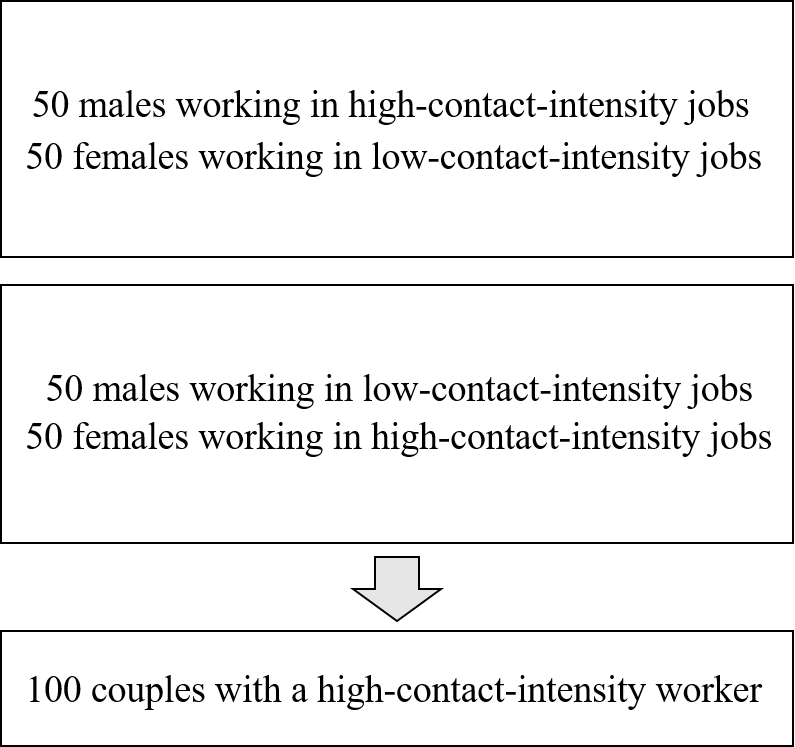}
		\subcaption{Perfect negative sorting}
		\label{fig: figure_2_ci}
	\end{subfigure}
	\caption{Spousal sorting by occupation contact intensity: An illustrative example}
	\label{fig: figure_ci}
\end{figure}

Workers whose occupations require high contact intensity at the workplace face higher risk of being infected, compared to those who work in low physical proximity to the other individuals.\footnote{~I consider contact intensity at the workplace, rather than contact intensity \textit{and} teleworkability, as a factor that accounts for contagion risk. A non-teleworkable occupation that does not require high contact intensity (e.g., operation of a machine) is not associated with higher risk of contagion at the workplace. Of course, it is possible to catch COVID-19 during commuting to work, and I control for this channel in my regressions.} Since the presence of the other family members creates the risk of intra-household contagion \citep{sun2020transmission}, then if at least one of the spouses work in a high CI occupation, the other family members are more exposed to COVID-19 risk. Consider an illustrative example described in Figure \ref{fig: figure_ci}. Suppose there are 100 males and 100 females that match into couples. Half of males work in high CI occupations, another half---in low CI occupations. Similarly, half of females work in high CI occupations, another half---in low CI occupations. First, consider perfect positive occupational sorting. In this case, as shown in Figure \ref{fig: figure_3_ci}, both spouses have either high CI or low CI jobs. The resulting distribution imply that there are 50 couples with at least one worker having a high CI job, and hence these 50 couples are more exposed to intra-household contagion risk. In particular, this risk is concentrated in high-contact-intensity couples. At the other extreme, as shown in Figure \ref{fig: figure_2_ci}, spouses have different jobs in terms of contact intensity (perfect negative sorting). The resulting distribution imply that there are 100 couples with at least one worker having a high CI job, and hence all 100 couples are exposed to intra-household contagion risk. To complement the discussion, in Figure \ref{fig: figure_2-5_ci}, I also show the case of zero sorting when males and females match at random. Under this scenario, 75 couples have at least one worker in a high CI job, and hence more exposed to within-household COVID-19 transmission. In a nutshell, this example shows that \textit{higher degree of positive occupational sorting} is associated with \textit{smaller number of individuals who are exposed to COVID-19 health risk}. Two populations characterized by similar distribution of males and females by occupation may demonstrate substantially different exposure to COVID-19 contagion risk depending on the patterns in spousal sorting. It is, therefore, an empirical question whether spousal occupational sorting is positive or negative in each particular population and, furthermore, how sizable it is. The scope of a question is significant because married couples constitute a substantial fraction of the U.S. population. According to the U.S. Bureau of the Census, in 2019 there were almost 62 million married couples. This accounts for about 48\% of all the U.S. households.

\begin{figure}[t!]
	\centering
	\includegraphics[width=.75\linewidth]{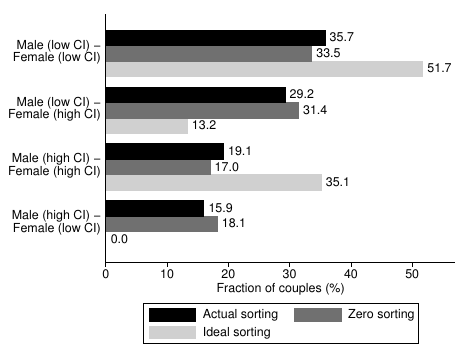}
	\caption{Distribution of dual-earner couples by occupation contact intensity in the United States: actual, zero, and ideal sorting (\%)}
	\label{fig: pp_sorting_us}
	\begin{spacing}{1.0}\justify\footnotesize{\textsc{Notes}: I use the 2015-2019 ACS individual data matched with the classification of occupations in terms of contact intensity from \cite{mongey2021which}. Low CI (``low contact intensity'') stands for occupations that do not require high contact intensity at the workplace. High CI (``high contact intensity'') stands for occupations that require high contact intensity at the workplace. Actual sorting corresponds to the distribution observed in the data. Zero sorting corresponds to random matching of spouses in terms of occupations. Ideal sorting is defined as the distribution where the fraction of couples with one spouse in a low CI job and another spouse in a high CI job is minimised.}\end{spacing}
\end{figure}

\subsection{Distribution of the U.S. Couples by Occupation Contact Intensity}\label{Distribution of the U.S. Couples by Occupation Contact Intensity}

I turn to characterization of spousal sorting by occupation in the United States. Figure \ref{fig: pp_sorting_us} reports the actual distribution of the U.S. dual-earner married couples by occupation contact intensity and compares it against two counterfactual distributions: zero sorting (or random matching) and ``ideal'' sorting. I define ideal sorting as the distribution where the fraction of ``mixed'' couples (one high CI and one low CI worker) is minimised or, in other words, it is maximum feasible positive sorting (i.e. the risk of intra-household contagion is minimized). The actual sorting (black bars) creates 64.2\% (29.2\% + 19.1\% + 15.9\%) couples with at least one spouse whose job requires high contact intensity at the workplace. These couples are exposed to greater intra-household contagion risk. Under zero sorting (dark grey bars), this fraction goes up to 66.5\% (31.4\% + 17.0\% + 18.1\%). Therefore, the existing occupational sorting in the U.S. couples creates a lower fraction of individuals who are exposed to intra-household contagion risk, compared to the case of zero sorting. Under ideal sorting (light grey bars), it falls down to 48.3\% (13.2\% + 35.1\%).

\begin{figure}[t!]
\centering
\includegraphics[width=0.95\linewidth]{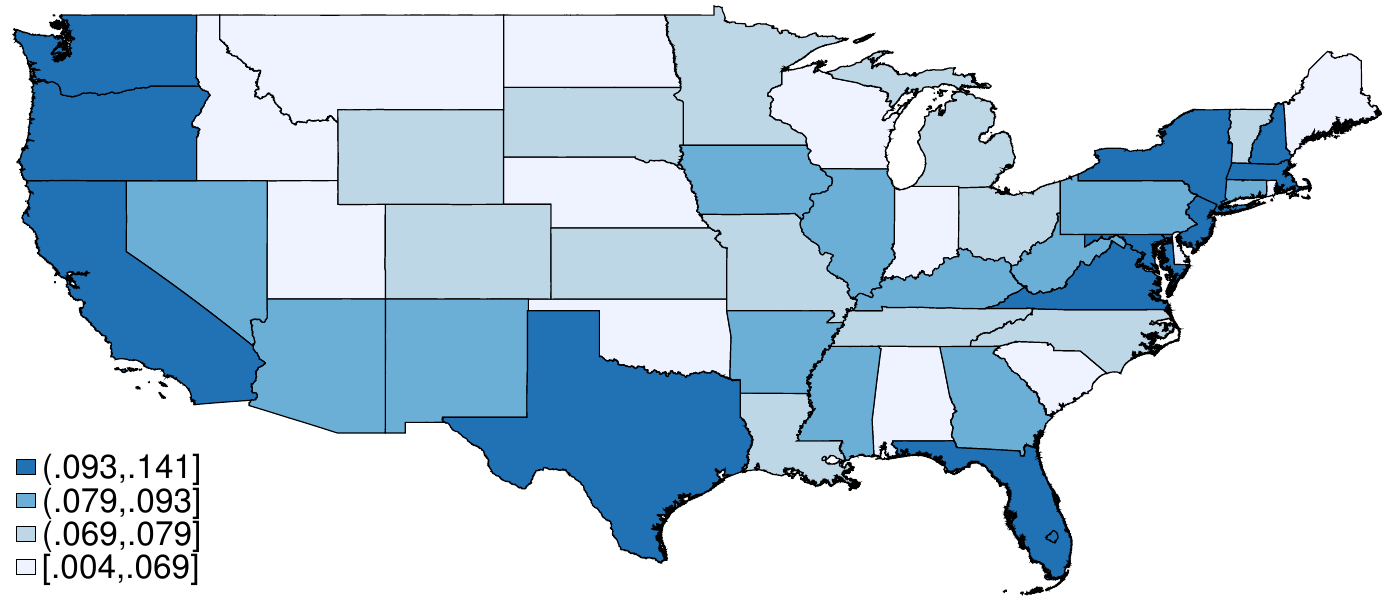}
\caption{Spousal (contact intensity) occupational sorting by state}
\label{fig: state_map}
\justify\footnotesize{\textsc{Notes}: I use the 2015-2019 ACS individual data matched with the classification of occupations in terms of contact intensity from \cite{mongey2021which}. The measure of spousal occupational sorting is calculated using \eqref{eq: occ_sorting_correlation}.}
\end{figure}

Using expression \eqref{eq: occ_sorting_correlation}, I obtain that the measure of spousal occupational sorting for the United States is equal to 0.091. As follows from Table \ref{tab: summary_statistics}, this aggregate statistic masks significant heterogeneity by state. In my regression analysis, I use the state-level variation in spousal sorting. Figure \ref{fig: state_map} shows the value of correlation \eqref{eq: occ_sorting_correlation} by state. First, no state is characterized by negative sorting. Second, from simple inspection, we observe the following spatial pattern: states with the highest levels of sorting are located on the West Coast, East Coast, and South (Texas).

\subsection{Spousal Occupational Sorting and COVID-19 Incidence}\label{Spousal Occupational Sorting and COVID-19 Incidence}

In this subsection, I present the main empirical results. In order to estimate the relationship between spousal occupational sorting and the incidence of COVID-19 and its evolution over time, I run the following regressions for every week $w(t)$:
\begin{equation}
\log \left( Total~Cases_{s w(t)}\right) = \alpha_t + \beta_{w(t)} Spousal~Occ.~Sorting_{s} + \boldsymbol{\delta}_{w(t)} \boldsymbol{X}_s + \varepsilon_{st}
\label{eq: regression_cases}
\end{equation}
\begin{equation}
\log \left( Total~Deaths_{s w(t)}\right) = \alpha_t + \beta_{w(t)} Spousal~Occ.~Sorting_{s} + \boldsymbol{\delta}_{w(t)} \boldsymbol{X}_s + \varepsilon_{st}
\label{eq: regression_deaths}
\end{equation}

\begin{spacing}{1}
	\begin{table}[t!]
		\begin{center}
		\caption{Cases by week} \label{tab: regressions_cases_week}
			\begin{adjustbox}{width=1\textwidth}
				\begin{tabular}{lcccccc}
					\hline \hline
					Dependent variable: & \multicolumn{6}{c}{Daily Log Cumulative Cases per 100000 Inhabitants} \\
					\cline{2-7}
					& \begin{tabular}[t]{@{}c@{}}Apr 1-7,\\2020\end{tabular} & \begin{tabular}[t]{@{}c@{}}Apr 8-14,\\2020\end{tabular} & \begin{tabular}[t]{@{}c@{}}Apr 15-21,\\2020\end{tabular} & \begin{tabular}[t]{@{}c@{}}Apr 22-28,\\2020\end{tabular} & \begin{tabular}[t]{@{}c@{}}Apr 29-May 5,\\2020\end{tabular} & \begin{tabular}[t]{@{}c@{}}May 6-12,\\2020\end{tabular} \\
					\hline
					Spousal occ. sorting (z-score) & \begin{tabular}[t]{@{}l@{}}-0.357$^{***}$\\(0.115)\end{tabular} & \begin{tabular}[t]{@{}l@{}}-0.281$^{**}$\\(0.119)\end{tabular} & \begin{tabular}[t]{@{}l@{}}-0.213$^{*}$\\(0.126)\end{tabular} & \begin{tabular}[t]{@{}l@{}}-0.153\\(0.112)\end{tabular} & \begin{tabular}[t]{@{}l@{}}-0.111\\(0.097)\end{tabular} & \begin{tabular}[t]{@{}l@{}}-0.079\\(0.093)\end{tabular} \\
					Log household size & \begin{tabular}[t]{@{}l@{}}2.309\\(6.382)\end{tabular} & \begin{tabular}[t]{@{}l@{}}2.081\\(5.973)\end{tabular} & \begin{tabular}[t]{@{}l@{}}2.955\\(5.731)\end{tabular} & \begin{tabular}[t]{@{}l@{}}1.583\\(5.165)\end{tabular} & \begin{tabular}[t]{@{}l@{}}-1.187\\(4.708)\end{tabular} & \begin{tabular}[t]{@{}l@{}}-2.555\\(4.424)\end{tabular} \\
					Share of married couples (\%) & \begin{tabular}[t]{@{}l@{}}-0.235$^{***}$\\(0.082)\end{tabular} & \begin{tabular}[t]{@{}l@{}}-0.225$^{***}$\\(0.081)\end{tabular} & \begin{tabular}[t]{@{}l@{}}-0.223$^{***}$\\(0.080)\end{tabular} & \begin{tabular}[t]{@{}l@{}}-0.194$^{**}$\\(0.075)\end{tabular} & \begin{tabular}[t]{@{}l@{}}-0.159$^{**}$\\(0.073)\end{tabular} & \begin{tabular}[t]{@{}l@{}}-0.136$^{*}$\\(0.070)\end{tabular} \\
					Share of males (\%) & \begin{tabular}[t]{@{}l@{}}1.992$^{***}$\\(0.579)\end{tabular} & \begin{tabular}[t]{@{}l@{}}1.642$^{***}$\\(0.511)\end{tabular} & \begin{tabular}[t]{@{}l@{}}1.223$^{**}$\\(0.476)\end{tabular} & \begin{tabular}[t]{@{}l@{}}0.904$^{**}$\\(0.434)\end{tabular} & \begin{tabular}[t]{@{}l@{}}0.676\\(0.405)\end{tabular} & \begin{tabular}[t]{@{}l@{}}0.457\\(0.395)\end{tabular} \\
					Neighborhood controls & Yes & Yes & Yes & Yes & Yes & Yes \\
					Demographic controls & Yes & Yes & Yes & Yes & Yes & Yes \\
					Occupation controls & Yes & Yes & Yes & Yes & Yes & Yes \\
					Day fixed effects & Yes & Yes & Yes & Yes & Yes & Yes \\
					\hline
					$R^2$ & 0.78 & 0.79 & 0.80 & 0.83 & 0.87 & 0.88 \\
					Observations & 357 & 357 & 357 & 357 & 357 & 357 \\
					\hline \hline
				    & \begin{tabular}[t]{@{}c@{}}May 13-19,\\2020\end{tabular} & \begin{tabular}[t]{@{}c@{}}May 20-26,\\2020\end{tabular} & \begin{tabular}[t]{@{}c@{}}May 27-Jun 2,\\2020\end{tabular} & \begin{tabular}[t]{@{}c@{}}Jun 3-9,\\2020\end{tabular} & \begin{tabular}[t]{@{}c@{}}Jun 10-16,\\2020\end{tabular} & \begin{tabular}[t]{@{}c@{}}Jun 17-23,\\2020\end{tabular} \\
					\hline
					Spousal occ. sorting (z-score) & \begin{tabular}[t]{@{}l@{}}-0.062\\(0.092)\end{tabular} & \begin{tabular}[t]{@{}l@{}}-0.056\\(0.091)\end{tabular} & \begin{tabular}[t]{@{}l@{}}-0.043\\(0.090)\end{tabular} & \begin{tabular}[t]{@{}l@{}}-0.023\\(0.087)\end{tabular} & \begin{tabular}[t]{@{}l@{}}-0.000\\(0.085)\end{tabular} & \begin{tabular}[t]{@{}l@{}}0.021\\(0.079)\end{tabular} \\
					Log household size & \begin{tabular}[t]{@{}l@{}}-2.749\\(4.255)\end{tabular} & \begin{tabular}[t]{@{}l@{}}-2.996\\(4.153)\end{tabular} & \begin{tabular}[t]{@{}l@{}}-3.401\\(4.041)\end{tabular} & \begin{tabular}[t]{@{}l@{}}-2.744\\(3.843)\end{tabular} & \begin{tabular}[t]{@{}l@{}}-1.776\\(3.633)\end{tabular} & \begin{tabular}[t]{@{}l@{}}-0.611\\(3.378)\end{tabular} \\
					Share of married couples (\%) & \begin{tabular}[t]{@{}l@{}}-0.119$^{*}$\\(0.068)\end{tabular} & \begin{tabular}[t]{@{}l@{}}-0.099\\(0.065)\end{tabular} & \begin{tabular}[t]{@{}l@{}}-0.078\\(0.061)\end{tabular} & \begin{tabular}[t]{@{}l@{}}-0.074\\(0.055)\end{tabular} & \begin{tabular}[t]{@{}l@{}}-0.070\\(0.050)\end{tabular} & \begin{tabular}[t]{@{}l@{}}-0.066\\(0.045)\end{tabular} \\
					Share of males (\%) & \begin{tabular}[t]{@{}l@{}}0.263\\(0.386)\end{tabular} & \begin{tabular}[t]{@{}l@{}}0.096\\(0.372)\end{tabular} & \begin{tabular}[t]{@{}l@{}}0.004\\(0.360)\end{tabular} & \begin{tabular}[t]{@{}l@{}}-0.081\\(0.341)\end{tabular} & \begin{tabular}[t]{@{}l@{}}-0.171\\(0.320)\end{tabular} & \begin{tabular}[t]{@{}l@{}}-0.240\\(0.304)\end{tabular} \\
					Neighborhood controls & Yes & Yes & Yes & Yes & Yes & Yes \\
					Demographic controls & Yes & Yes & Yes & Yes & Yes & Yes \\
					Occupation controls & Yes & Yes & Yes & Yes & Yes & Yes \\
					Day fixed effects & Yes & Yes & Yes & Yes & Yes & Yes \\
					\hline
					$R^2$ & 0.89 & 0.90 & 0.91 & 0.91 & 0.92 & 0.92 \\
					Observations & 357 & 357 & 357 & 357 & 357 & 357 \\
					\hline \hline
				\end{tabular}
			\end{adjustbox}
			\justify\footnotesize{\textsc{Notes}: All columns show versions of OLS \eqref{eq: regression_cases} for different weeks. The dependent variables is log of daily cumulative cases per 100000 inhabitants up to date. All controls are described in the text. Standard errors are clustered at the state level and shown in parentheses. Significance: * $p < 0.1$, ** $p < 0.05$, *** $p < 0.01$.}
		\end{center}
	\end{table}
\end{spacing}
\bigskip

Regressions \eqref{eq: regression_cases} and \eqref{eq: regression_deaths} differ only in the dependent variables. In \eqref{eq: regression_cases}, I use log daily cumulative cases per 100000 inhabitants in state $s$ pooled in week $w(t)$. In \eqref{eq: regression_deaths}, I use log daily cumulative deaths per 100000 inhabitants in state $s$ pooled in week $w(t)$. My approach allows the coefficients to be time-varying, so that we can study the interaction of demographic and socioeconomic state characteristics with the evolution of COVID-19 pandemic. The main variable of interest, $Spousal~Occ.~Sorting_{s}$, is given by the standardized correlation \eqref{eq: occ_sorting_correlation} for state $s$, i.e. it has zero mean and a standard deviation of one. Hence, a one standard deviation increase in the measure of spousal occupational sorting is associated with $\beta_{w(t)}$ log points (or $100 \times \left( \exp \left( \beta_{w(t)} \right) - 1 \right)\%$) increase in cumulative cases (or deaths) per 100000 inhabitants in week $w(t)$. Furthermore, my regressions include day fixed effects $\alpha_t$ that control for common factors across all the states, and a battery of control variables that are widely considered as potential factors of COVID-19 spread. Namely, I control for the log average household size, share of married couples, share of males, shares of different age groups (20-39, 40-59, and above 60), shares of Black, Hispanic, and Asian population, share of people that do not have health insurance, log median income, log population density, average commute time to work, share of people who use public transportation, and, finally, employment of working-age population by occupation groups in state $s$. Finally, in all the regressions I cluster the standard errors at the state level.

\begin{figure}[t!]
	\centering
	\begin{subfigure}{.49\textwidth}
		\centering
		\includegraphics[width=\linewidth]{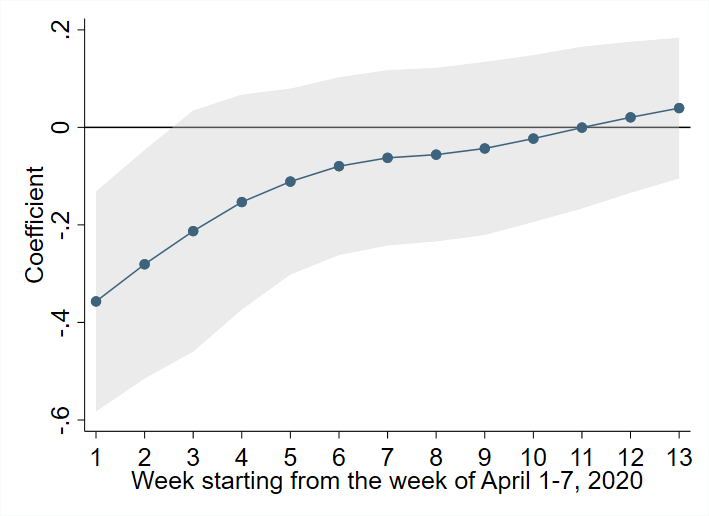}
		\label{fig: corr_pp_z_cases}
		\subcaption{Effect on the number of cases per 100000}
	\end{subfigure}%
	\begin{subfigure}{.49\textwidth}
		\centering
		\includegraphics[width=\linewidth]{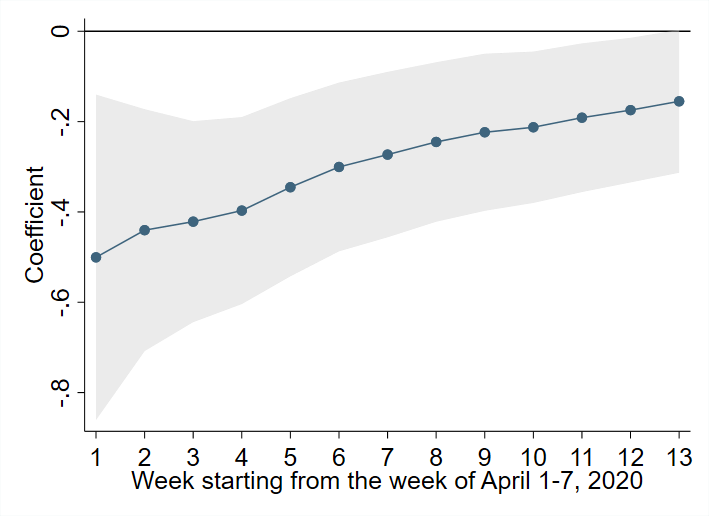}
		\label{fig: corr_pp_z_deaths}
		\subcaption{Effect on the number of deaths per 100000}
	\end{subfigure}
	\caption{Weekly evolution of coefficient of spousal occupational sorting}
	\label{fig: coeff_corr_pp_z}
	\begin{spacing}{1.0}\justify\footnotesize{\textsc{Notes}: Coefficients are from OLS \eqref{eq: regression_cases} (left panel) and \eqref{eq: regression_deaths} (right panel) for different weeks. The shaded area represents the 95\% confidence interval.}\end{spacing}
\end{figure}

\begin{figure}[b!]
	\centering
	\begin{subfigure}{.33\textwidth}
		\centering
		\includegraphics[width=\linewidth]{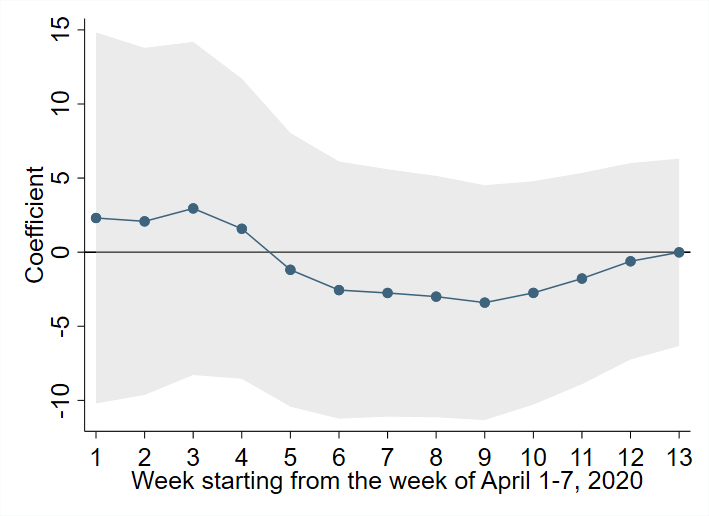}
		\label{fig: log_hh_size_cases}
		\subcaption{Log household size (cases)}
	\end{subfigure}%
	\begin{subfigure}{.33\textwidth}
		\centering
		\includegraphics[width=\linewidth]{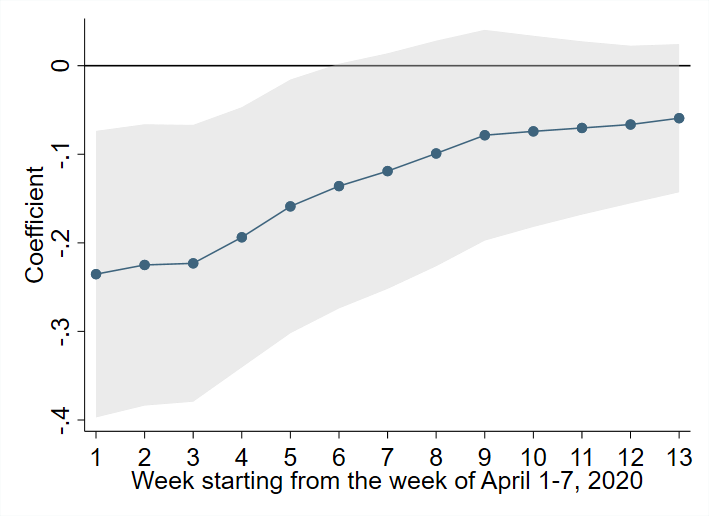}
		\label{fig: couples_share_cases}
		\subcaption{Share of couples (cases)}
	\end{subfigure}%
	\begin{subfigure}{.33\textwidth}
		\centering
		\includegraphics[width=\linewidth]{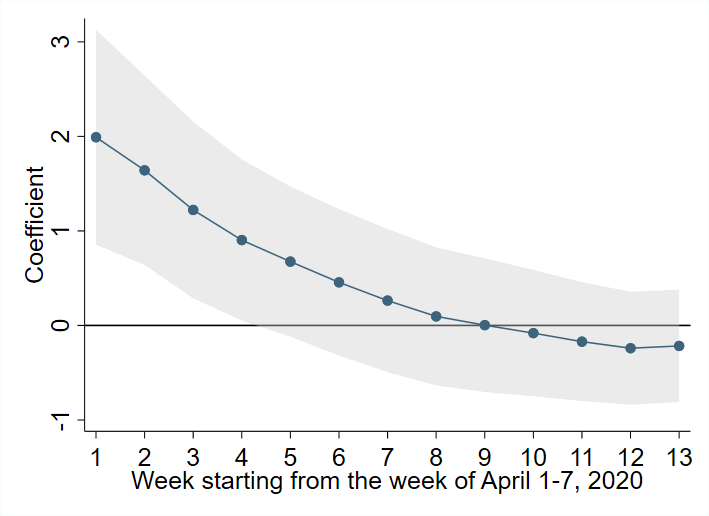}
		\label{fig: male_share_cases}
		\subcaption{Share of males (cases)}
	\end{subfigure}
	\begin{subfigure}{.33\textwidth}
		\centering
		\includegraphics[width=\linewidth]{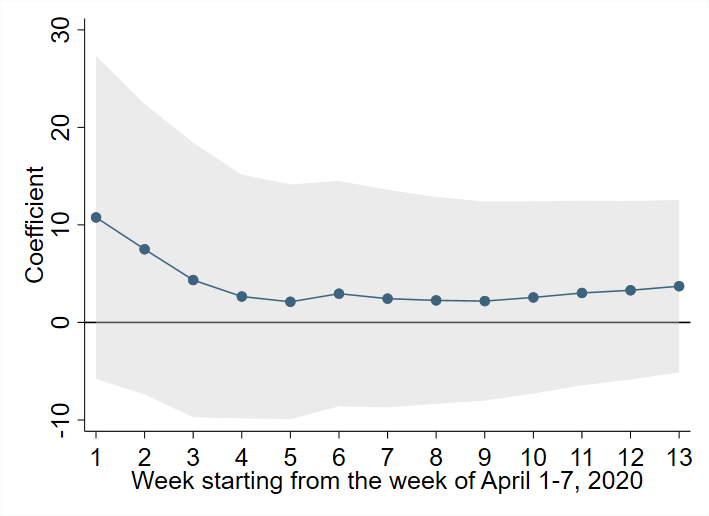}
		\label{fig: log_hh_size_deaths}
		\subcaption{Log household size (deaths)}
	\end{subfigure}%
	\begin{subfigure}{.33\textwidth}
		\centering
		\includegraphics[width=\linewidth]{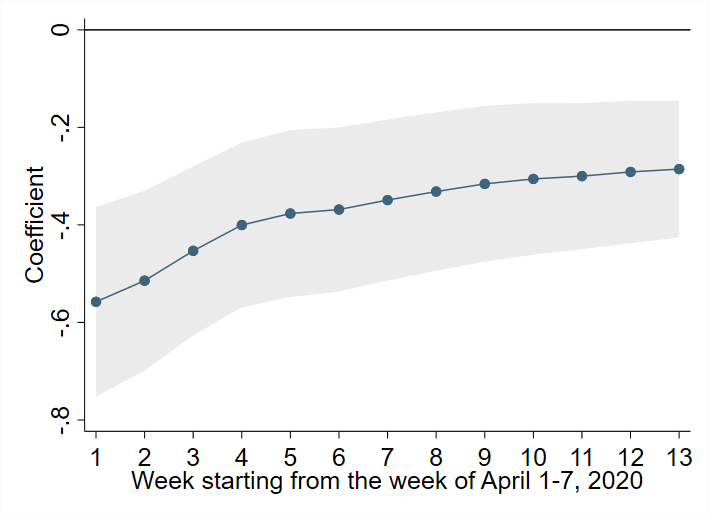}
		\label{fig: couples_share_deaths}
		\subcaption{Share of couples (deaths)}
	\end{subfigure}%
	\begin{subfigure}{.33\textwidth}
		\centering
		\includegraphics[width=\linewidth]{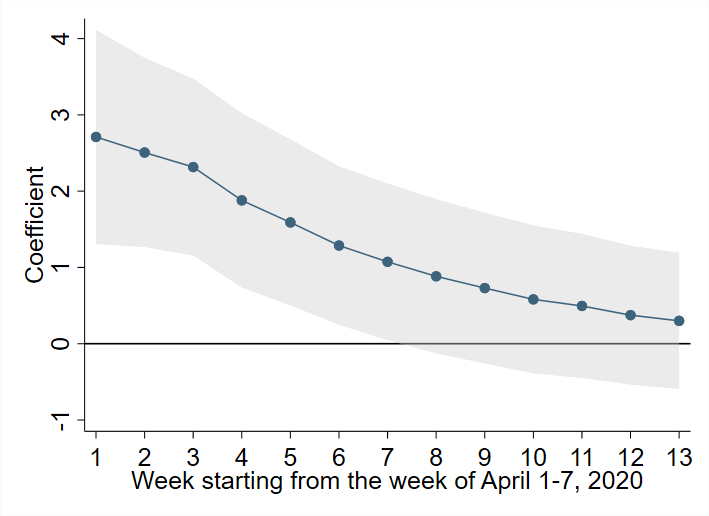}
		\label{fig: male_share_deaths}
		\subcaption{Share of males (deaths)}
	\end{subfigure}
	\caption{Weekly evolution of coefficients of household variables}
	\label{fig: coeff_hh_variables}
	\begin{spacing}{1.0}\justify\footnotesize{\textsc{Notes}: Coefficients are from OLS \eqref{eq: regression_cases} (upper panel) and \eqref{eq: regression_deaths} (lower panel) for different weeks. The shaded area represents the 95\% confidence interval.}\end{spacing}
\end{figure}

Table \ref{tab: regressions_cases_week} shows the results for COVID-19 cases over twelve weeks between April 1-7, 2020 and June 17-23, 2020. In particular, I report the coefficients for the household characteristics. Additionally, I plot the evolution of these coefficients and their 95\% confidence intervals on the left panel of Figure \ref{fig: coeff_corr_pp_z} and the top panel of Figure \ref{fig: coeff_hh_variables}. The first finding is that spousal occupational sorting is a statistically significant factor over the early weeks of the pandemic. For example, for the week of April 1-7, a one standard deviation increase in the measure of spousal occupational sorting\footnote{~For example, this corresponds to the difference between Oregon ($\rho = 0.101$) and New York ($\rho = 0.123$).} is associated with a decrease in the cumulative number of cases per 100000 inhabitants by 0.357 log points (or $100 \times \left( \exp \left( -0.357 \right) - 1 \right) = 30\%$). Given that for the week of April 1-7, the average cumulative number of cases over states was 72.4 per 100000 inhabitants, a one standard deviation increase in spousal occupational sorting would lower this to 50.9 per 100000 inhabitants.\footnote{~It is worth noting that the coefficients may be biased due to the measurement error. One source of the measurement error comes from the fact that a sizeable share of workers lost their jobs in the first months of the pandemic, and hence the pre-pandemic occupation shares and the measure of spousal occupational sorting differ from the actual ones. To minimize this measurement error, I conduct my analysis on the first weeks of the pandemic. Furthermore, the number of COVID-19 cases and deaths as well as the variables from the ACS data (that uses a 5\% sample) may be measured with error.} The magnitude of the coefficient decreases over time, and by the third week, April 15-21, a one standard deviation increase in the measure of spousal sorting is correlated with a decrease in the cumulative number of cases per 100000 inhabitants by 0.213 log points. Starting from the week of April 22-28, the coefficient is no longer statistically significant.

In addition, Table \ref{tab: regressions_cases_week} and Figure \ref{fig: coeff_hh_variables} report that the share of married couples is negatively correlated, and the share of males is positively correlated with the number of COVID-19 cases. For example, for the week of April 1-7, a one percentage point increase in the share of married couples is associated with a 15.2 decrease ($72.4 \times \left( \exp \left( -0.235 \right) - 1 \right)$) in the cumulative number of cases per 100000 inhabitants. I also find that household size does not have a significant effect on the number of cases. Furthermore, Figure \ref{fig: coeff_cases_other} reports the evolution of coefficients of the other variables over time. Consistent with the findings from \cite{almagro2020nyc} and \cite{glaeser2021much} who analyze the data for New York City, such factors as population density, use of public transportation, and the share of Hispanic population are correlated with higher spread of COVID-19. Furthermore, higher employment shares in Non-Essential Professional, Non-Essential Service, Law Enforcement, and Transportation occupation categories are associated with higher number of COVID-19 cases.

\begin{spacing}{1}
	\begin{table}[t!]
		\begin{center}
			\caption{Deaths by week}
			\label{tab: regressions_deaths_week}
			\begin{adjustbox}{width=1\textwidth}
				\begin{tabular}{lcccccc}
					\hline \hline
					Dependent variable: & \multicolumn{6}{c}{Daily Log Cumulative Cases per 100000 Inhabitants} \\
					\cline{2-7}
					& \begin{tabular}[t]{@{}c@{}}Apr 1-7,\\2020\end{tabular} & \begin{tabular}[t]{@{}c@{}}Apr 8-14,\\2020\end{tabular} & \begin{tabular}[t]{@{}c@{}}Apr 15-21,\\2020\end{tabular} & \begin{tabular}[t]{@{}c@{}}Apr 22-28,\\2020\end{tabular} & \begin{tabular}[t]{@{}c@{}}Apr 29-May 5,\\2020\end{tabular} & \begin{tabular}[t]{@{}c@{}}May 6-12,\\2020\end{tabular} \\
					\hline
					Spousal occ. sorting (z-score) & \begin{tabular}[t]{@{}l@{}}-0.500$^{***}$\\(0.184)\end{tabular} & \begin{tabular}[t]{@{}l@{}}-0.440$^{***}$\\(0.137)\end{tabular} & \begin{tabular}[t]{@{}l@{}}-0.421$^{***}$\\(0.114)\end{tabular} & \begin{tabular}[t]{@{}l@{}}-0.397$^{***}$\\(0.106)\end{tabular} & \begin{tabular}[t]{@{}l@{}}-0.345$^{***}$\\(0.101)\end{tabular} & \begin{tabular}[t]{@{}l@{}}-0.300$^{***}$\\(0.095)\end{tabular} \\
					Log household size & \begin{tabular}[t]{@{}l@{}}10.772\\(8.448)\end{tabular} & \begin{tabular}[t]{@{}l@{}}7.507\\(7.593)\end{tabular} & \begin{tabular}[t]{@{}l@{}}4.351\\(7.176)\end{tabular} & \begin{tabular}[t]{@{}l@{}}2.663\\(6.373)\end{tabular} & \begin{tabular}[t]{@{}l@{}}2.121\\(6.138)\end{tabular} & \begin{tabular}[t]{@{}l@{}}2.955\\(5.890)\end{tabular} \\
					Share of married couples (\%) & \begin{tabular}[t]{@{}l@{}}-0.558$^{***}$\\(0.099)\end{tabular} & \begin{tabular}[t]{@{}l@{}}-0.514$^{***}$\\(0.094)\end{tabular} & \begin{tabular}[t]{@{}l@{}}-0.453$^{***}$\\(0.088)\end{tabular} & \begin{tabular}[t]{@{}l@{}}-0.400$^{***}$\\(0.086)\end{tabular} & \begin{tabular}[t]{@{}l@{}}-0.377$^{***}$\\(0.087)\end{tabular} & \begin{tabular}[t]{@{}l@{}}-0.368$^{***}$\\(0.086)\end{tabular} \\
					Share of males (\%) & \begin{tabular}[t]{@{}l@{}}2.712$^{***}$\\(0.716)\end{tabular} & \begin{tabular}[t]{@{}l@{}}2.508$^{***}$\\(0.632)\end{tabular} & \begin{tabular}[t]{@{}l@{}}2.316$^{***}$\\(0.591)\end{tabular} & \begin{tabular}[t]{@{}l@{}}1.880$^{***}$\\(0.583)\end{tabular} & \begin{tabular}[t]{@{}l@{}}1.591$^{***}$\\(0.555)\end{tabular} & \begin{tabular}[t]{@{}l@{}}1.288$^{**}$\\(0.530)\end{tabular} \\
					Neighborhood controls & Yes & Yes & Yes & Yes & Yes & Yes \\
					Demographic controls & Yes & Yes & Yes & Yes & Yes & Yes \\
					Occupation controls & Yes & Yes & Yes & Yes & Yes & Yes \\
					Day fixed effects & Yes & Yes & Yes & Yes & Yes & Yes \\
					\hline
					$R^2$ & 0.81 & 0.84 & 0.87 & 0.88 & 0.89 & 0.90 \\
					Observations & 350 & 352 & 357 & 357 & 357 & 357 \\
					\hline \hline
					& \begin{tabular}[t]{@{}c@{}}May 13-19,\\2020\end{tabular} & \begin{tabular}[t]{@{}c@{}}May 20-26,\\2020\end{tabular} & \begin{tabular}[t]{@{}c@{}}May 27-Jun 2,\\2020\end{tabular} & \begin{tabular}[t]{@{}c@{}}Jun 3-9,\\2020\end{tabular} & \begin{tabular}[t]{@{}c@{}}Jun 10-16,\\2020\end{tabular} & \begin{tabular}[t]{@{}c@{}}Jun 17-23,\\2020\end{tabular} \\
					\hline
					Spousal occ. sorting (z-score) & \begin{tabular}[t]{@{}l@{}}-0.273$^{***}$\\(0.093)\end{tabular} & \begin{tabular}[t]{@{}l@{}}-0.245$^{***}$\\(0.090)\end{tabular} & \begin{tabular}[t]{@{}l@{}}-0.224$^{**}$\\(0.089)\end{tabular} & \begin{tabular}[t]{@{}l@{}}-0.212$^{**}$\\(0.085)\end{tabular} & \begin{tabular}[t]{@{}l@{}}-0.191$^{**}$\\(0.084)\end{tabular} & \begin{tabular}[t]{@{}l@{}}-0.174$^{**}$\\(0.082)\end{tabular} \\
					Log household size & \begin{tabular}[t]{@{}l@{}}2.444\\(5.687)\end{tabular} & \begin{tabular}[t]{@{}l@{}}2.266\\(5.407)\end{tabular} & \begin{tabular}[t]{@{}l@{}}2.196\\(5.207)\end{tabular} & \begin{tabular}[t]{@{}l@{}}2.567\\(5.021)\end{tabular} & \begin{tabular}[t]{@{}l@{}}3.029\\(4.826)\end{tabular} & \begin{tabular}[t]{@{}l@{}}3.303\\(4.665)\end{tabular} \\
					Share of married couples (\%) & \begin{tabular}[t]{@{}l@{}}-0.349$^{***}$\\(0.084)\end{tabular} & \begin{tabular}[t]{@{}l@{}}-0.331$^{***}$\\(0.083)\end{tabular} & \begin{tabular}[t]{@{}l@{}}-0.316$^{***}$\\(0.081)\end{tabular} & \begin{tabular}[t]{@{}l@{}}-0.306$^{***}$\\(0.079)\end{tabular} & \begin{tabular}[t]{@{}l@{}}-0.300$^{***}$\\(0.076)\end{tabular} & \begin{tabular}[t]{@{}l@{}}-0.291$^{***}$\\(0.074)\end{tabular} \\
					Share of males (\%) & \begin{tabular}[t]{@{}l@{}}1.074$^{**}$\\(0.524)\end{tabular} & \begin{tabular}[t]{@{}l@{}}0.884$^{*}$\\(0.516)\end{tabular} & \begin{tabular}[t]{@{}l@{}}0.730\\(0.503)\end{tabular} & \begin{tabular}[t]{@{}l@{}}0.582\\(0.495)\end{tabular} & \begin{tabular}[t]{@{}l@{}}0.496\\(0.482)\end{tabular} & \begin{tabular}[t]{@{}l@{}}0.375\\(0.465)\end{tabular} \\
					Neighborhood controls & Yes & Yes & Yes & Yes & Yes & Yes \\
					Demographic controls & Yes & Yes & Yes & Yes & Yes & Yes \\
					Occupation controls & Yes & Yes & Yes & Yes & Yes & Yes \\
					Day fixed effects & Yes & Yes & Yes & Yes & Yes & Yes \\
					\hline
					$R^2$ & 0.91 & 0.91 & 0.91 & 0.92 & 0.92 & 0.92 \\
					Observations & 357 & 357 & 357 & 357 & 357 & 357 \\
					\hline \hline
				\end{tabular}
			\end{adjustbox}
			\justify\footnotesize{\textsc{Notes}: All columns show versions of OLS \eqref{eq: regression_deaths} for different weeks. The dependent variables is log of daily cumulative deaths per 100000 inhabitants up to date. All controls are described in the text. Standard errors are clustered at the state level and shown in parentheses. Significance: * $p < 0.1$, ** $p < 0.05$, *** $p < 0.01$.}
		\end{center}
	\end{table}
\end{spacing}
\medskip

Next, Table \ref{tab: regressions_deaths_week} reports the results for COVID-19 deaths over twelve weeks between April 1-7, 2020 and June 17-23, 2020. The right panel of Figure \ref{fig: coeff_corr_pp_z} and the bottom panel of Figure \ref{fig: coeff_hh_variables} plot the evolution of coefficients over time. The first lesson from these results is that spousal occupational sorting is a statistically significant factor until the week of June 17-23. Speaking about magnitudes, for the week of April 1-7, a one standard deviation increase in the measure of spousal occupational sorting is associated with a decline in the cumulative number of deaths per 100000 inhabitants by 0.5 log points (or $100 \times \left( \exp \left( -0.5 \right) - 1 \right) = 39.3\%$). Given that for the week of April 1-7, the average cumulative number of deaths over states was 2.2 per 100000 inhabitants, a one standard deviation increase in spousal occupational sorting would lower this to 1.3 per 100000. Similarly to the regressions with the number of cases per capita as a dependent variable, the magnitude of the coefficient declines over time. For the week of April 15-21, a one standard deviation increase in the measure of spousal occupational sorting is associated with a decrease in the cumulative number of deaths per 100000 inhabitants by 0.421 log points, for the week of May 13-19---by 0.273 log points, and for the week of June 10-16---by 0.191 log points. Overall, my results confirm the hypothesis that sorting of spouses by occupation contact intensity has a significant effect on the prevalence of COVID-19 in the United States.

\newpage
Furthermore, Table \ref{tab: regressions_deaths_week} and Figure \ref{fig: coeff_hh_variables} show that the share of married couples is negatively correlated and the share of males is positively correlated with the cumulative number of deaths. The finding about married couples is consistent with \cite{drefahl2020population} who, using the Swedish data, document that unmarried individuals have a higher risk of death from COVID-19. This pattern may be partly explained by selection: people with worse health are less attractive to potential partners, and hence have less chances to get married. The finding about males confirms the results of \cite{bwire2020coronavirus}, \cite{drefahl2020population}, and \cite{peckham2020male} who show that male patients have higher odds of death. Moreover, the average household size again does not have a significant effect. Similarly to the coefficients for spousal sorting, a general observation from Tables \ref{tab: regressions_cases_week}-\ref{tab: regressions_deaths_week} and Figure \ref{fig: coeff_hh_variables} is that the coefficients of the other household characteristics decrease in magnitude over time. Next, Figure \ref{fig: coeff_deaths_other} plots the evolution of coefficients of the other variables over time. We observe the positive significant effects of population density, use of public transportation, and employment shares in Non-Essential Professional, Non-Essential Service, and Industrial and Construction categories on the number of COVID-19 deaths. In turn, the share of young population (aged 20-39) and employment shares in Law and Related and Essential Service categories are associated with lower number of deaths. Similarly to the household characteristics, most of the coefficients demonstrate declining magnitude over time.

\section{Conclusion}\label{Conclusion}

In this paper, I present evidence that the degree of spousal occupational sorting is an important factor for explaining disparities in the incidence of COVID-19 across the United States during the early stages of the pandemic. In particular, I test the hypothesis that a higher degree of positive sorting is associated with a lower incidence of COVID-19. To address this question, I construct the state-level measure of spousal sorting using the ACS data and merge it with the daily data on COVID-19 cases and deaths provided by the JHU CSSE.

My first step is to document the distribution of the U.S. dual-earner couples by occupation contact intensity. I show that spousal sorting creates 64.2\% families that are exposed to greater intra-household contagion risk because at least one spouse has a high contact intensity job. Furthermore, I uncover substantial heterogeneity in the degree of spousal sorting by state.

Next, using variation in the degree of spousal sorting by state, I estimate its effects on the number of COVID-19 cases and deaths per 100000 inhabitants. In particular, I run the regressions for each week over the period between April 1, 2020 and July 1, 2020, hence allowing the coefficients to be time-varying. The estimates imply that, for the week of April 1-7, a one standard deviation increase in the measure of spousal sorting is associated with a 30\% reduction in the number of cases per 100000 and a 39.3\% decline in the number of deaths per 100000. Furthermore, I find that the coefficients decline in magnitude over time and eventually lose statistical significance. Beyond that, I show that the share of males is positively correlated and the share of married couples is negatively correlated with the number of cases and deaths per capita.

My findings emphasize the importance of interaction between the nature of work and spousal sorting for the incidence of COVID-19 and suggest several policy implications. On the one hand, targeting workers in jobs that require high contact intensity with testing and vaccination, and providing them with protective equipment, would indirectly mitigate the intra-household contagion channel as their family members would be less likely exposed to the risk of being infected. On the other hand, the other policy tools may directly address the problem of within-household transmission.

In this paper, I focus on spousal occupational sorting and hence limit my analysis to dual-earner couples. An important direction for future research is to explore the role of the other family-related factors in the incidence of COVID-19. Furthermore, my analysis can be extended by considering the economic outcomes for couples with different exposure to contact intensity at work. In the previous version of this paper, I document a negative relationship between the total household income and the share of couples where at least one spouse should work in high contact intensity. Hence the COVID-19 pandemic likely exposes poorer households to greater health risks. Moreover, while I briefly discuss several policy implications, a more careful and formal study of optimal policies is necessary. \cite{baqaee2020reopening} is an example of a quantitative paper that studies the economic reopening using the data on contact intensity and teleworkability by sector. Furthermore, my results emphasize the importance of accounting for contacts between individuals, both at work and within a household, for designing the optimal policies \citep{azzimonti2020pandemic}. Last but not least, my analysis can be extended using the data from the other countries. These are fruitful avenues for future research.

\section*{Declarations}\label{Declarations}

\subsection*{Conflicts of Interests}\label{Conflicts of Interests}

The author has no relevant financial or non-financial interests to disclose.

\subsection*{Funding}\label{Funding}

The author did not receive support from any organization for the submitted work.

\begin{spacing}{1}
\bibliographystyle{ecta}
\bibliography{draft_june2021}
\end{spacing}

\newpage

\setcounter{equation}{0}
\setcounter{figure}{0}
\setcounter{table}{0}
\renewcommand\theequation{A.\arabic{equation}}
\renewcommand\thefigure{A.\arabic{figure}}
\renewcommand\thetable{A.\arabic{table}}

\thispagestyle{empty}

\begin{spacing}{1}
\begin{table}[t!]
\footnotesize
\begin{center}
\caption{Spousal occupational sorting by state}
\label{tab: spousal_sorting_states}
\begin{tabular}{lccccccc}
\hline \hline
State & $\pi_{hh}$ & $\pi_{hl}$ & $\pi_{lh}$ & $\pi_{ll}$ & $q_h^m$ & $q_h^m$ & $\rho$ \\
\hline
Alabama	& 0.187 & 0.161 & 0.319 & 0.334 & 0.347 & 0.506 & 0.046 \\
Alaska & 0.230 & 0.181 & 0.265 & 0.324 & 0.411 & 0.495 & 0.107 \\
Arizona	& 0.204 & 0.172 & 0.285 & 0.339 & 0.376 & 0.489 & 0.084 \\
Arkansas & 0.214 & 0.159 & 0.306 & 0.321 & 0.373 & 0.520 & 0.083 \\
California & 0.184 & 0.161 & 0.267 & 0.388 & 0.345 & 0.451 & 0.119 \\
Colorado & 0.174 & 0.163 & 0.287 & 0.376 & 0.337 & 0.460 & 0.079 \\
Connecticut & 0.180 & 0.161 & 0.284 & 0.375 & 0.341 & 0.464 & 0.093 \\
Delaware & 0.185 & 0.180 & 0.279 & 0.356 & 0.365 & 0.464 & 0.066 \\
District of Columbia & 0.079 & 0.139 & 0.169 & 0.613 & 0.218 & 0.248 & 0.141 \\
Florida & 0.215 & 0.177 & 0.271 & 0.337 & 0.392 & 0.486 & 0.100 \\
Georgia & 0.184 & 0.155 & 0.300 & 0.362 & 0.338 & 0.484 & 0.084 \\
Hawaii & 0.254 & 0.201 & 0.255 & 0.291 & 0.454 & 0.509 & 0.091 \\
Idaho & 0.192 & 0.154 & 0.317 & 0.338 & 0.345 & 0.509 & 0.067 \\
Illinois & 0.184 & 0.156 & 0.296 & 0.364 & 0.339 & 0.480 & 0.088 \\
Indiana & 0.191 & 0.153 & 0.319 & 0.337 & 0.344 & 0.510 & 0.066 \\
Iowa & 0.182 & 0.149 & 0.304 & 0.365 & 0.331 & 0.486 & 0.088 \\
Kansas & 0.186 & 0.145 & 0.324 & 0.345 & 0.331 & 0.510 & 0.073 \\
Kentucky & 0.209 & 0.156 & 0.311 & 0.324 & 0.365 & 0.520 & 0.080 \\
Louisiana & 0.211 & 0.168 & 0.295 & 0.326 & 0.379 & 0.505 & 0.079 \\
Maine & 0.203 & 0.177 & 0.294 & 0.326 & 0.380 & 0.497 & 0.060 \\
Maryland & 0.165 & 0.157 & 0.267 & 0.411 & 0.322 & 0.432 & 0.112 \\
Massachusetts & 0.174 & 0.152 & 0.282 & 0.392 & 0.326 & 0.456 & 0.109 \\
Michigan & 0.184 & 0.153 & 0.311 & 0.352 & 0.337 & 0.495 & 0.072 \\
Minnesota & 0.169 & 0.153 & 0.305 & 0.373 & 0.322 & 0.474 & 0.072 \\
Mississippi & 0.225 & 0.158 & 0.313 & 0.305 & 0.383 & 0.538 & 0.080 \\
Missouri & 0.196 & 0.169 & 0.293 & 0.341 & 0.366 & 0.489 & 0.072 \\
Montana & 0.188 & 0.177 & 0.299 & 0.336 & 0.365 & 0.487 & 0.043 \\
Nebraska & 0.171 & 0.150 & 0.316 & 0.362 & 0.322 & 0.488 & 0.062 \\
Nevada & 0.267 &0.205 & 0.254 & 0.274 & 0.472 & 0.521 & 0.084 \\
New Hampshire & 0.185 & 0.149 & 0.301 & 0.364 & 0.334 & 0.487 & 0.097 \\
New Jersey & 0.179 & 0.156 & 0.282 & 0.383 & 0.335 & 0.461 & 0.105 \\
New Mexico & 0.217 & 0.174 & 0.288 & 0.321 & 0.391 & 0.505 & 0.081 \\
New York & 0.224 & 0.169 & 0.269 & 0.338 & 0.393 & 0.493 & 0.123 \\
North Carolina & 0.185 & 0.157 & 0.301 & 0.357 & 0.342 & 0.486 & 0.079 \\
North Dakota & 0.165 & 0.164 & 0.335 & 0.336 & 0.329 & 0.500 & 0.004 \\
Ohio & 0.187 & 0.152 & 0.314 & 0.347 & 0.339 & 0.501 & 0.072 \\
Oklahoma & 0.195 & 0.161 & 0.314 & 0.331 & 0.355 & 0.508 & 0.058 \\
Oregon & 0.191 & 0.162 & 0.281 & 0.365 & 0.353 & 0.472 & 0.101 \\
Pennsylvania & 0.191 & 0.156 & 0.300 & 0.353 & 0.347 & 0.491 & 0.086 \\
Rhode Island & 0.188 & 0.174 & 0.287 & 0.351 & 0.362 & 0.475 & 0.067 \\
South Carolina & 0.192 & 0.160 & 0.309 & 0.339 & 0.352 & 0.501 & 0.066 \\
South Dakota & 0.180 & 0.148 & 0.313 & 0.359 & 0.328 & 0.494 & 0.078 \\
Tennessee & 0.204 & 0.161 & 0.304 & 0.331 & 0.365 & 0.508 & 0.076 \\
Texas & 0.194 & 0.154 & 0.295 & 0.356 & 0.348 & 0.489 & 0.099 \\
Utah & 0.179 & 0.146 & 0.335 & 0.341 & 0.325 & 0.513 & 0.052 \\
Vermont & 0.186 & 0.168 & 0.288 & 0.357 & 0.354 & 0.475 & 0.076 \\
Virginia & 0.171 & 0.151 & 0.286 & 0.393 & 0.321 & 0.457 & 0.103 \\
Washington & 0.182 & 0.150 & 0.293 & 0.376 & 0.331 & 0.474 & 0.104 \\
West Virginia &	0.235 & 0.154 & 0.319 & 0.292 & 0.389 & 0.554 & 0.080 \\
Wisconsin &	0.177 & 0.157 & 0.305 & 0.361 & 0.334 & 0.482 & 0.069 \\
Wyoming & 0.203 & 0.164 & 0.298 & 0.335 & 0.367 & 0.501 & 0.079 \\
\hline \hline
\end{tabular}
\justify\footnotesize{\textsc{Notes}: I use the 2015-2019 ACS individual data matched with the classification of occupations in terms of contact intensity from \cite{mongey2021which}. Columns 2-5 report the distribution of couples by occupations in terms of contact intensity. Columns 6-7 report the fraction of males and females working in high contact intensity occupations. The last column reports the measure of spousal occupational sorting calculated using \eqref{eq: occ_sorting_correlation}.}
\end{center}
\end{table}
\end{spacing}

\newpage

\begin{sidewaysfigure}
\begin{figure}[H]
	\centering
	\begin{subfigure}{.166\textwidth}
		\centering
		\includegraphics[width=\linewidth]{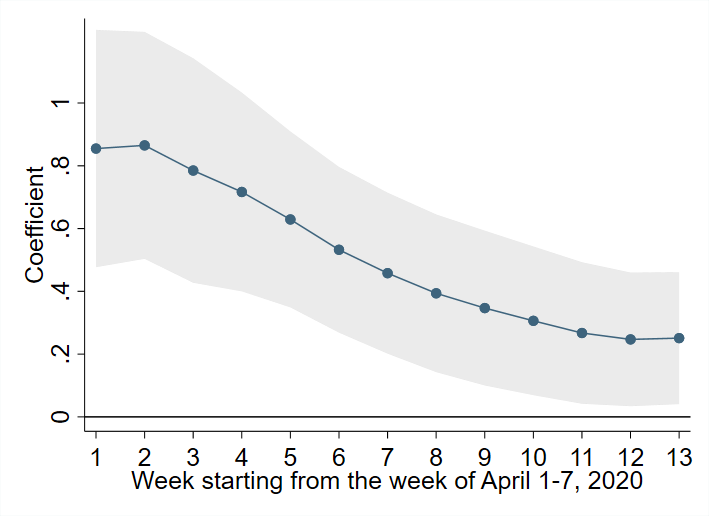}
		\label{fig: log_density_cases}
		\subcaption{Log density}
	\end{subfigure}%
	\begin{subfigure}{.166\textwidth}
		\centering
		\includegraphics[width=\linewidth]{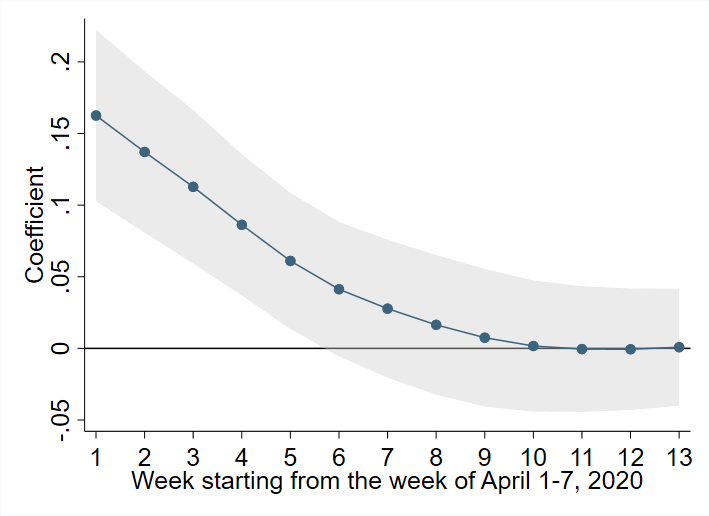}
		\label{fig: public_transport_cases}
		\subcaption{Public transportation}
	\end{subfigure}%
	\begin{subfigure}{.166\textwidth}
		\centering
		\includegraphics[width=\linewidth]{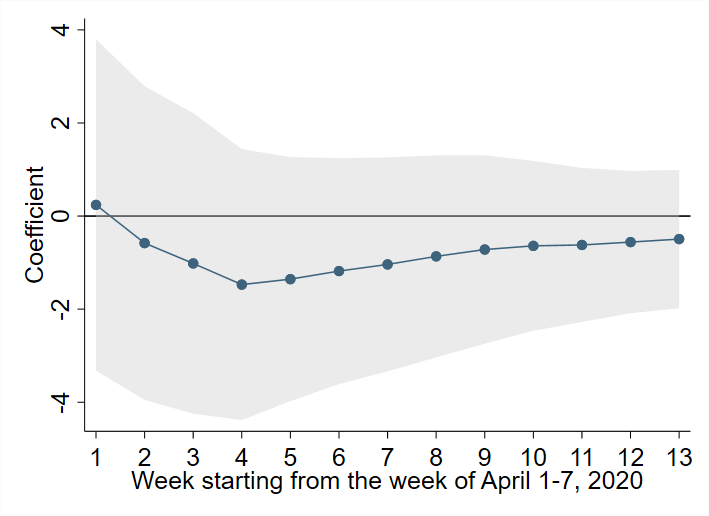}
		\label{fig: log_avg_time_cases}
		\subcaption{Commute time}
	\end{subfigure}%
	\begin{subfigure}{.166\textwidth}
		\centering
		\includegraphics[width=\linewidth]{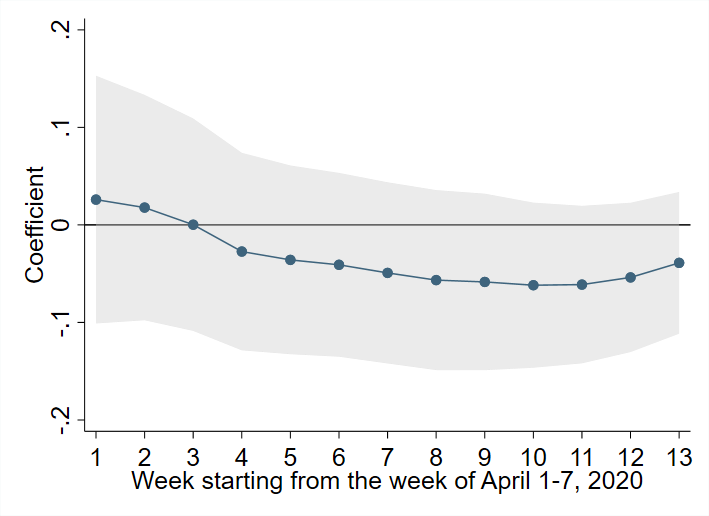}
		\label{fig: total_no_hins_cases}
		\subcaption{No health insurance}
	\end{subfigure}%
	\begin{subfigure}{.166\textwidth}
		\centering
		\includegraphics[width=\linewidth]{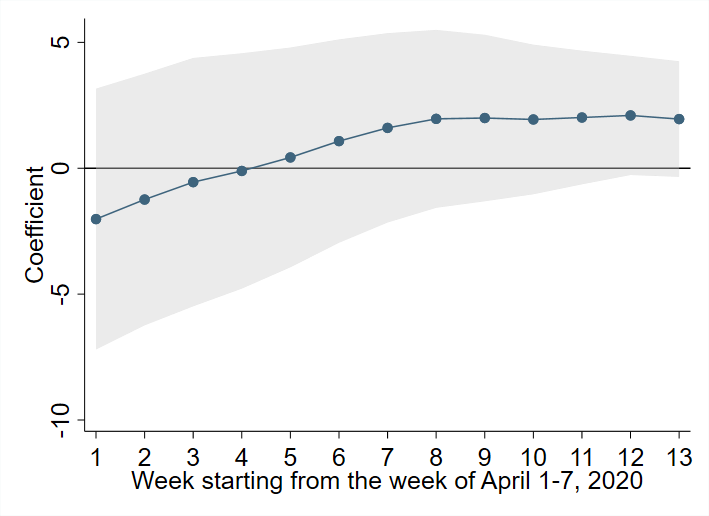}
		\label{fig: log_med_income_cases}
		\subcaption{Log median income}
	\end{subfigure}%
	\begin{subfigure}{.166\textwidth}
		\centering
		\includegraphics[width=\linewidth]{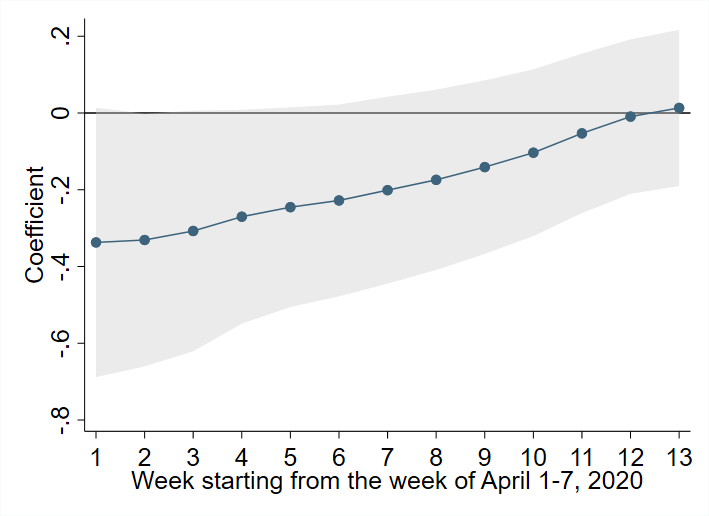}
		\label{fig: age_20_39_cases}
		\subcaption{Share of age 20-39}
	\end{subfigure}
	\begin{subfigure}{.166\textwidth}
		\centering
		\includegraphics[width=\linewidth]{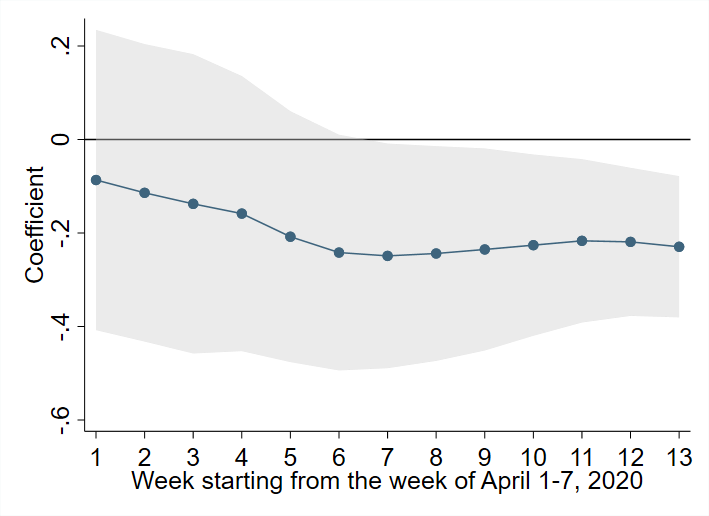}
		\label{fig: age_40_59_cases}
		\subcaption{Share of age 40-59}
	\end{subfigure}%
	\begin{subfigure}{.166\textwidth}
		\centering
		\includegraphics[width=\linewidth]{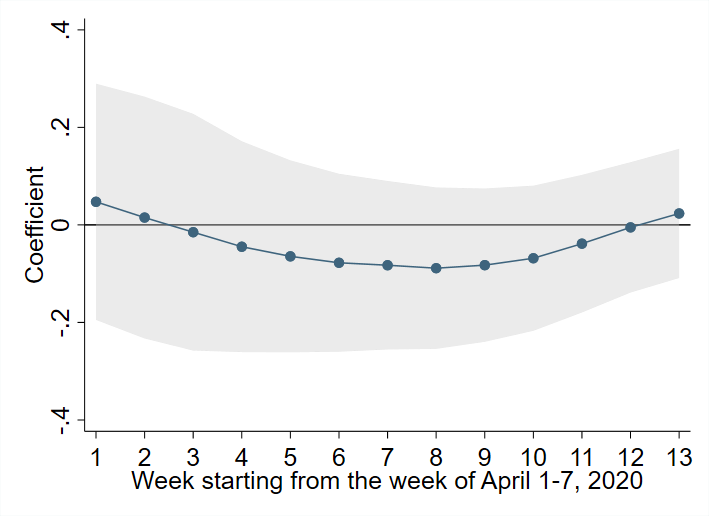}
		\label{fig: age_60_plus_cases}
		\subcaption{Share of age 60+}
	\end{subfigure}%
	\begin{subfigure}{.166\textwidth}
		\centering
		\includegraphics[width=\linewidth]{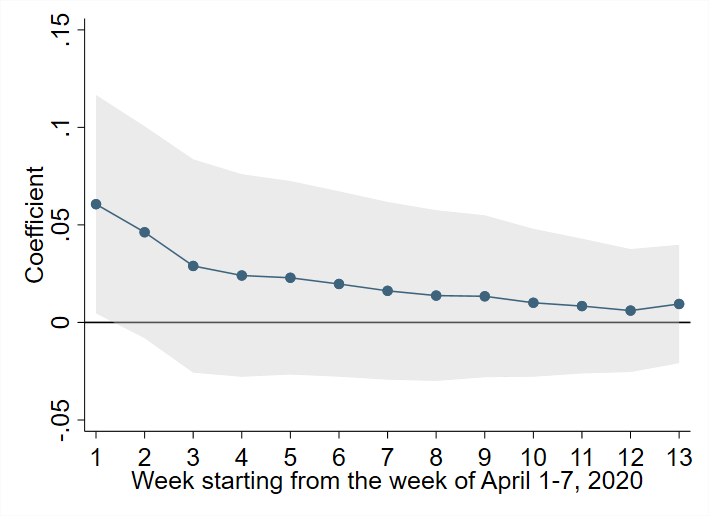}
		\label{fig: black_share_cases}
		\subcaption{Share of Black}
	\end{subfigure}%
	\begin{subfigure}{.166\textwidth}
		\centering
		\includegraphics[width=\linewidth]{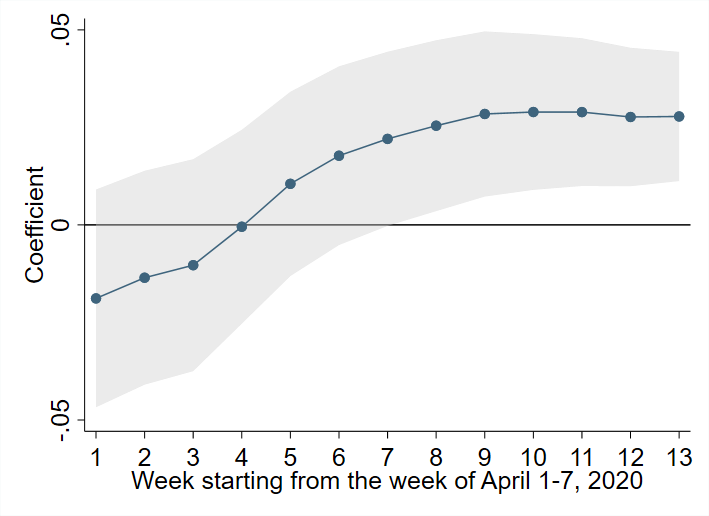}
		\label{fig: hispanic_share_cases}
		\subcaption{Share of Hispanic}
	\end{subfigure}%
	\begin{subfigure}{.166\textwidth}
		\centering
		\includegraphics[width=\linewidth]{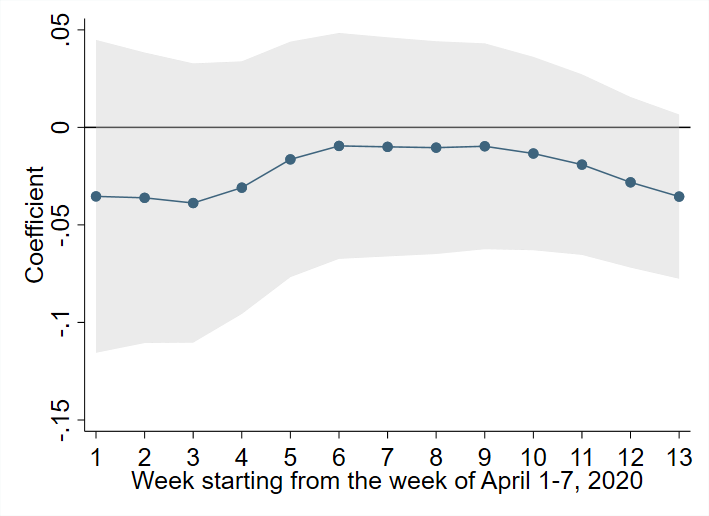}
		\label{fig: asian_share_cases}
		\subcaption{Share of Asian}
	\end{subfigure}%
	\begin{subfigure}{.166\textwidth}
		\centering
		\includegraphics[width=\linewidth]{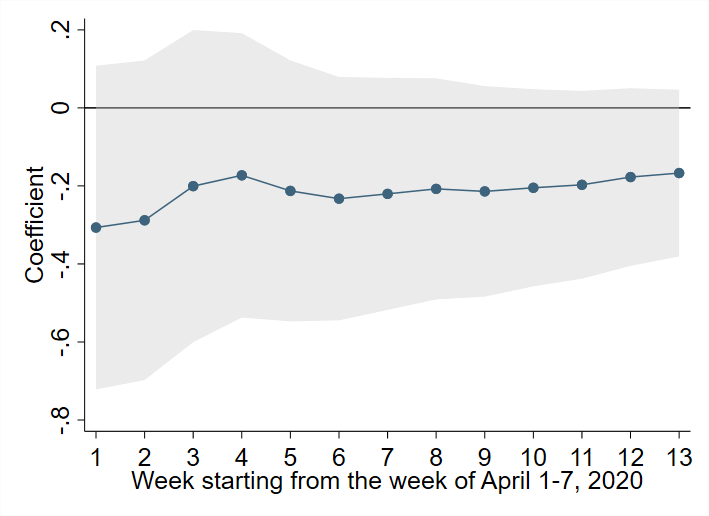}
		\label{fig: occ_1_cases}
		\subcaption{Ess-Professional}
	\end{subfigure}
	\begin{subfigure}{.166\textwidth}
		\centering
		\includegraphics[width=\linewidth]{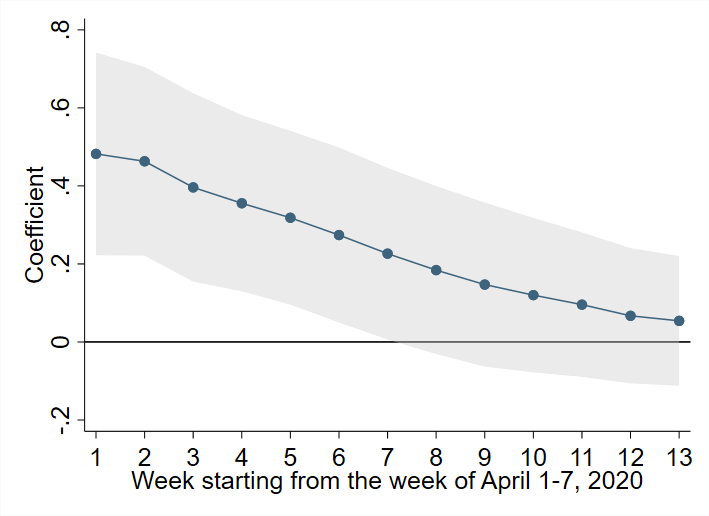}
		\label{fig: occ_2_cases}
		\subcaption{Non-Ess-Prof}
	\end{subfigure}%
	\begin{subfigure}{.166\textwidth}
		\centering
		\includegraphics[width=\linewidth]{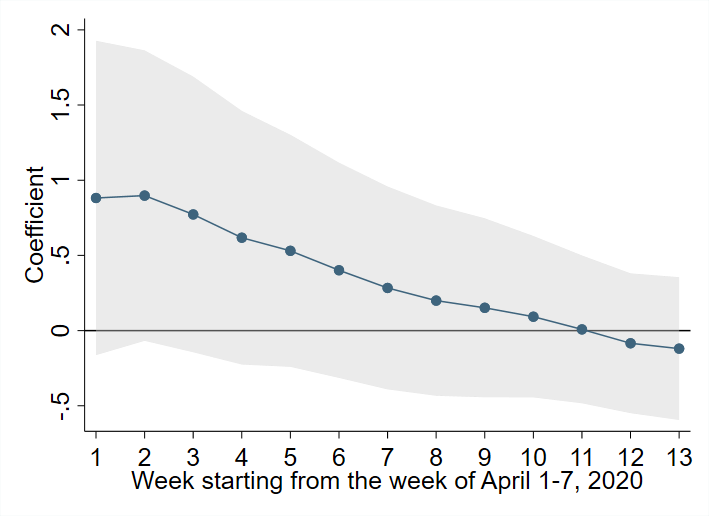}
		\label{fig: occ_3_cases}
		\subcaption{Science}
	\end{subfigure}%
	\begin{subfigure}{.166\textwidth}
		\centering
		\includegraphics[width=\linewidth]{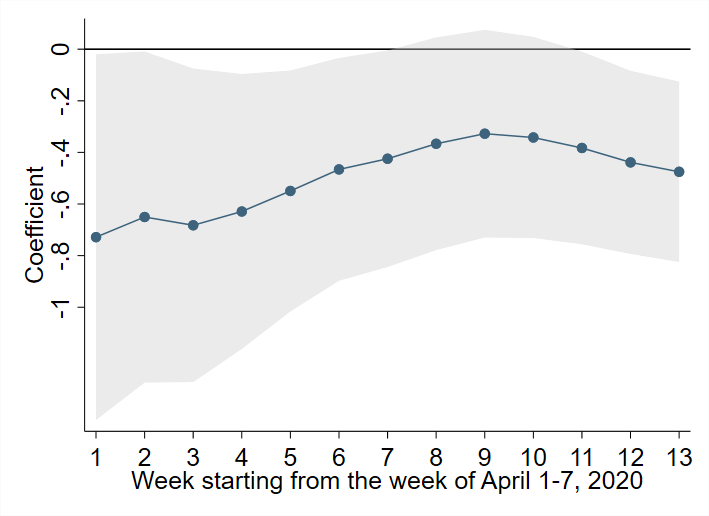}
		\label{fig: occ_4_cases}
		\subcaption{Law and Related}
	\end{subfigure}%
	\begin{subfigure}{.166\textwidth}
		\centering
		\includegraphics[width=\linewidth]{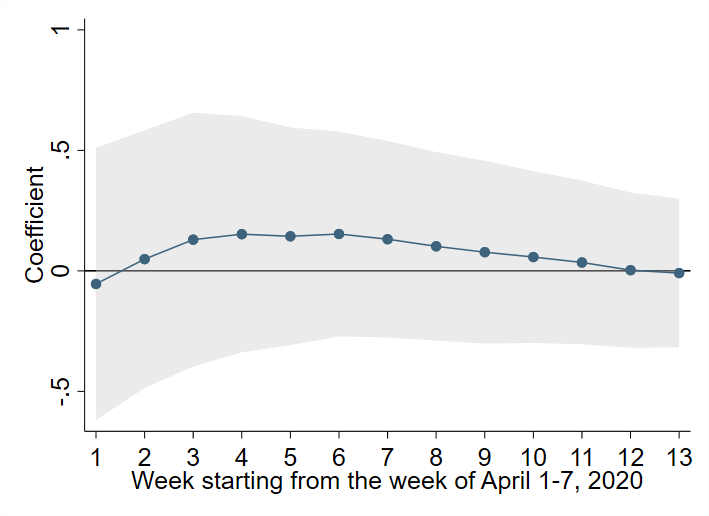}
		\label{fig: occ_5_cases}
		\subcaption{Health Practitioners}
	\end{subfigure}%
	\begin{subfigure}{.166\textwidth}
		\centering
		\includegraphics[width=\linewidth]{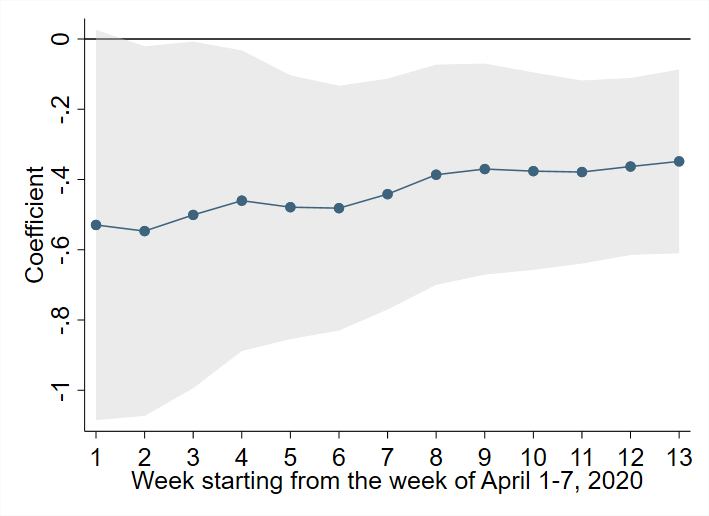}
		\label{fig: occ_6_cases}
		\subcaption{Other Health}
	\end{subfigure}%
	\begin{subfigure}{.166\textwidth}
		\centering
		\includegraphics[width=\linewidth]{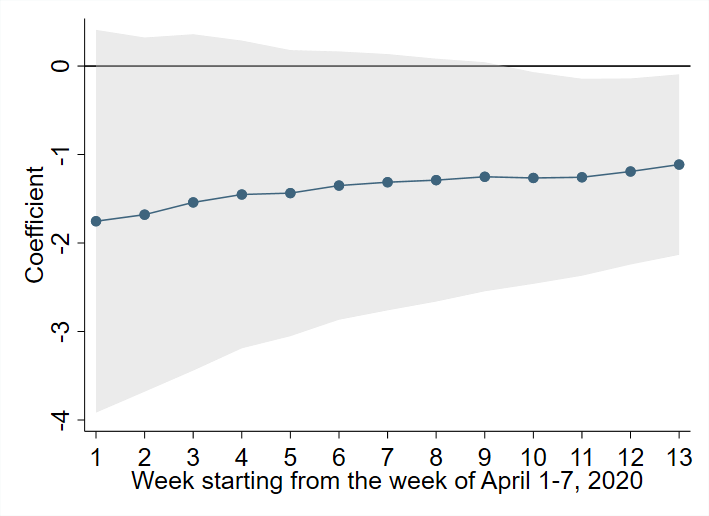}
		\label{fig: occ_7_cases}
		\subcaption{Firefighting}
	\end{subfigure}
	\begin{subfigure}{.166\textwidth}
		\centering
		\includegraphics[width=\linewidth]{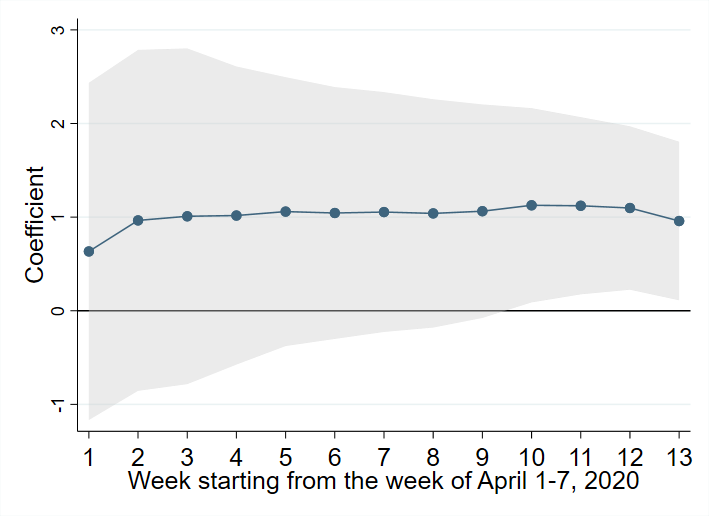}
		\label{fig: occ_8_cases}
		\subcaption{Law Enforcement}
	\end{subfigure}%
	\begin{subfigure}{.166\textwidth}
		\centering
		\includegraphics[width=\linewidth]{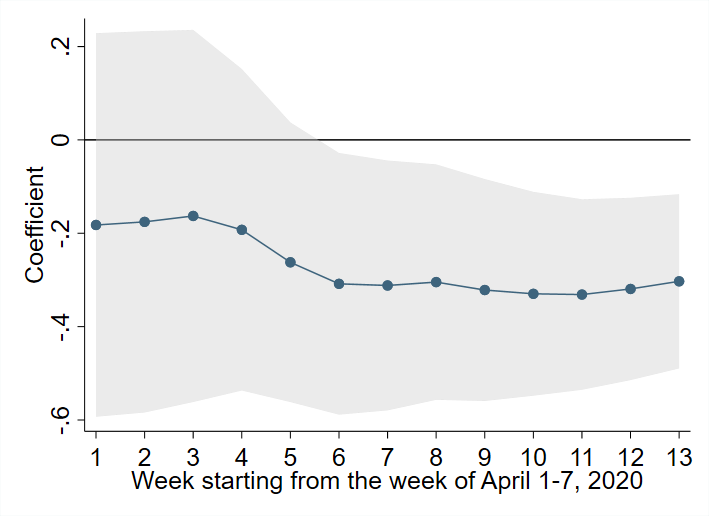}
		\label{fig: occ_9_cases}
		\subcaption{Essential-Service}
	\end{subfigure}%
	\begin{subfigure}{.166\textwidth}
		\centering
		\includegraphics[width=\linewidth]{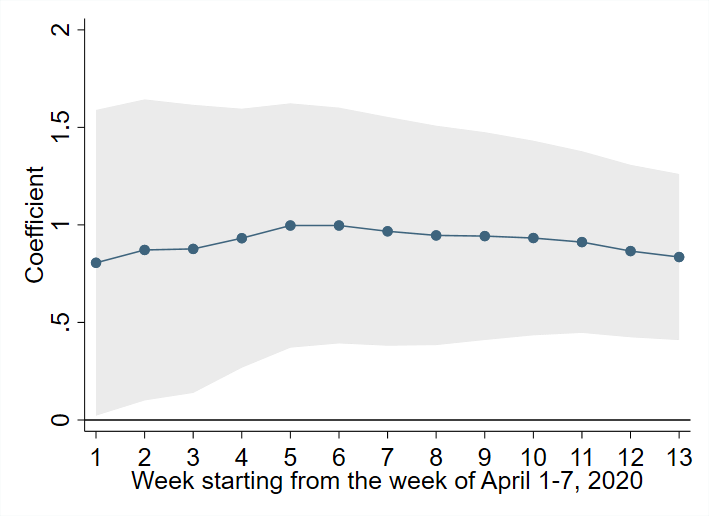}
		\label{fig: occ_10_cases}
		\subcaption{Non-Ess-Service}
	\end{subfigure}%
	\begin{subfigure}{.166\textwidth}
		\centering
		\includegraphics[width=\linewidth]{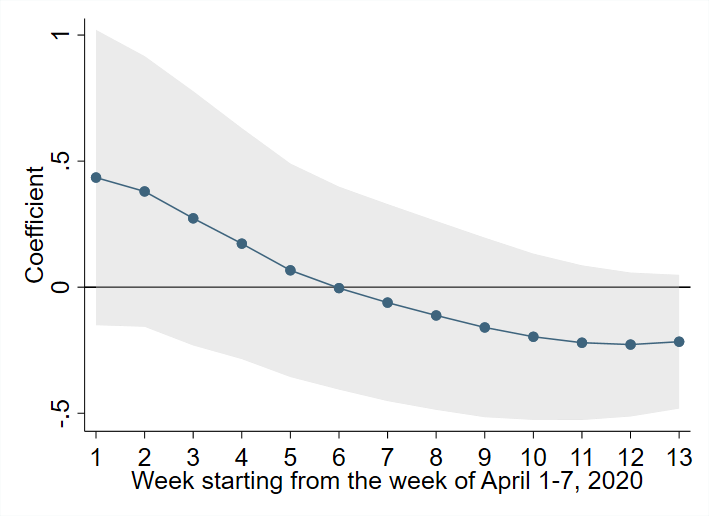}
		\label{fig: occ_11_cases}
		\subcaption{Industrial \& Constr.}
	\end{subfigure}%
	\begin{subfigure}{.166\textwidth}
		\centering
		\includegraphics[width=\linewidth]{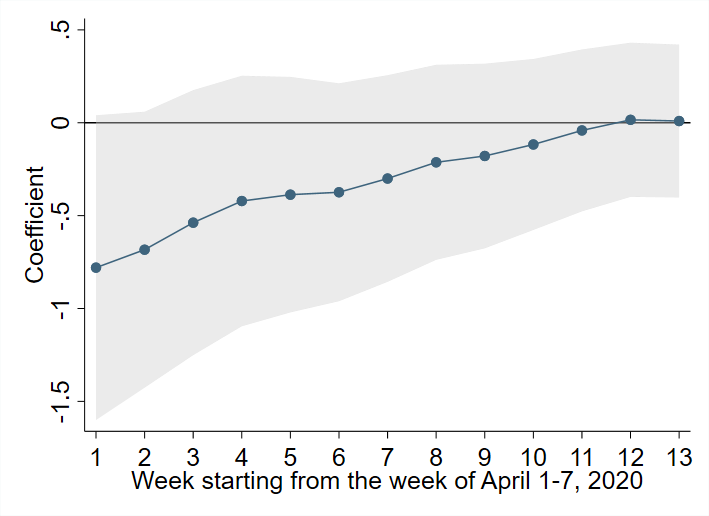}
		\label{fig: occ_12_cases}
		\subcaption{Essential-Technical}
	\end{subfigure}%
	\begin{subfigure}{.166\textwidth}
		\centering
		\includegraphics[width=\linewidth]{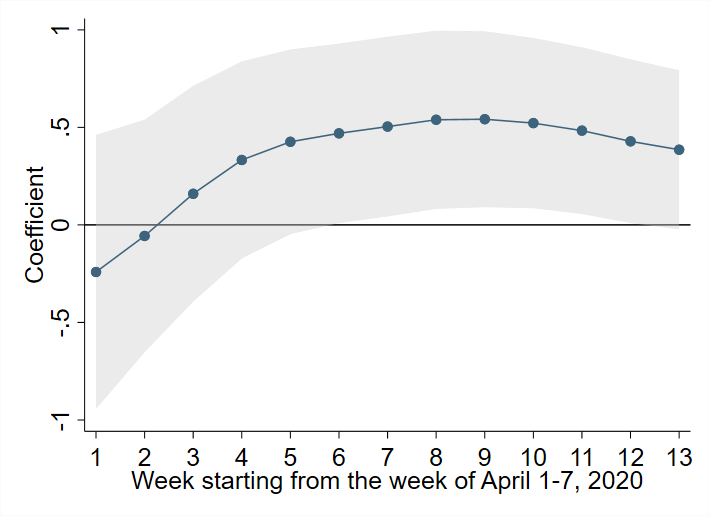}
		\label{fig: occ_13_cases}
		\subcaption{Transportation}
	\end{subfigure}%
	\caption{Weekly evolution of coefficients (dependent variable: log of daily cumulative cases per 100000 inhabitants)}
	\label{fig: coeff_cases_other}
	\begin{spacing}{1.0}\justify\footnotesize{\textsc{Notes}: The figure shows the impact of state characteristics on COVID-19 cases. Coefficients are from OLS regression \eqref{eq: regression_cases} for the weeks from April 1-7, 2020 to June 24-July 1, 2020. The shaded area represents the 95\% confidence interval.}\end{spacing}
\end{figure}
\end{sidewaysfigure}

\begin{sidewaysfigure}
\begin{figure}[H]
	\centering
	\begin{subfigure}{.166\textwidth}
		\centering
		\includegraphics[width=\linewidth]{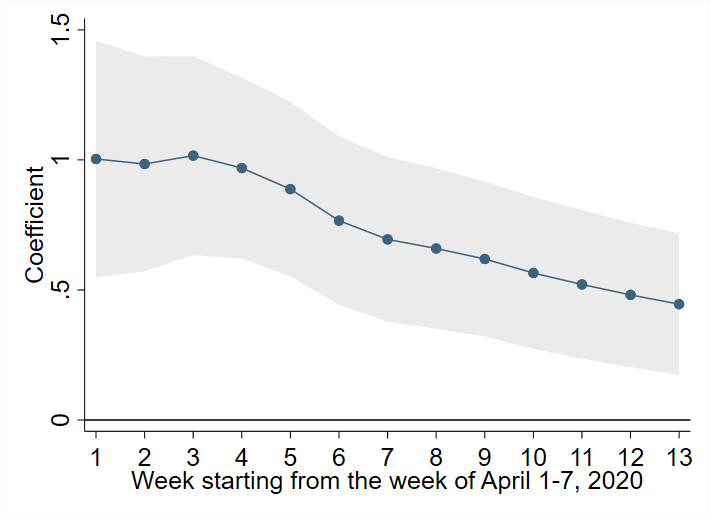}
		\label{fig: log_density_deaths}
		\subcaption{Log density}
	\end{subfigure}%
	\begin{subfigure}{.166\textwidth}
		\centering
		\includegraphics[width=\linewidth]{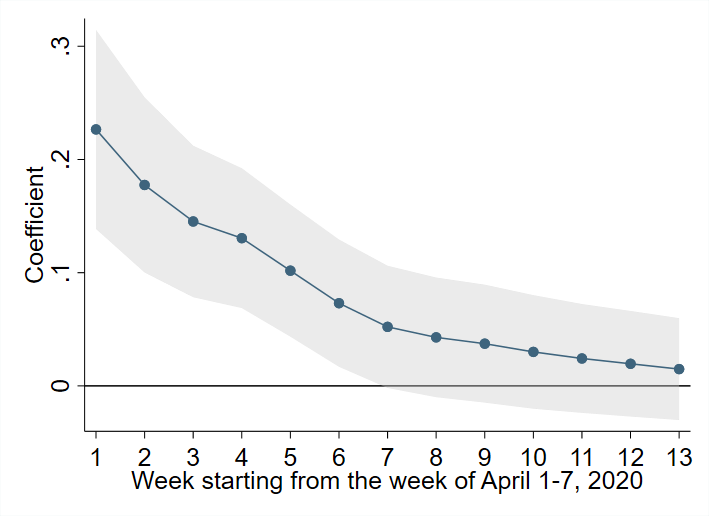}
		\label{fig: public_transport_deaths}
		\subcaption{Public transportation}
	\end{subfigure}%
	\begin{subfigure}{.166\textwidth}
		\centering
		\includegraphics[width=\linewidth]{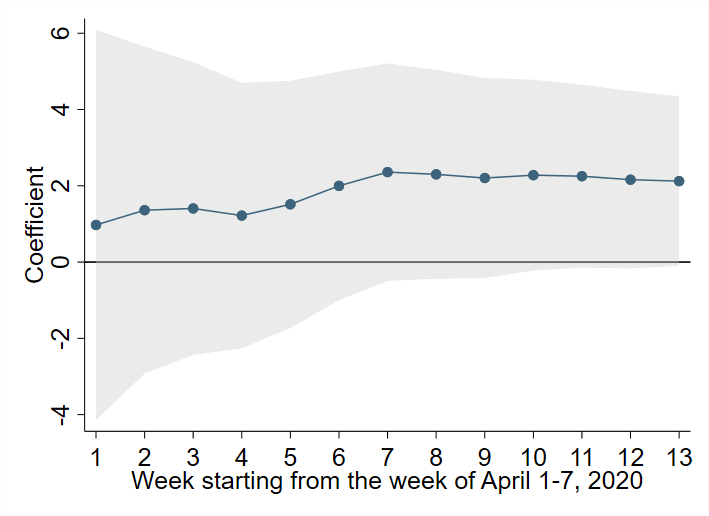}
		\label{fig: log_avg_time_deaths}
		\subcaption{Commute time}
	\end{subfigure}%
	\begin{subfigure}{.166\textwidth}
		\centering
		\includegraphics[width=\linewidth]{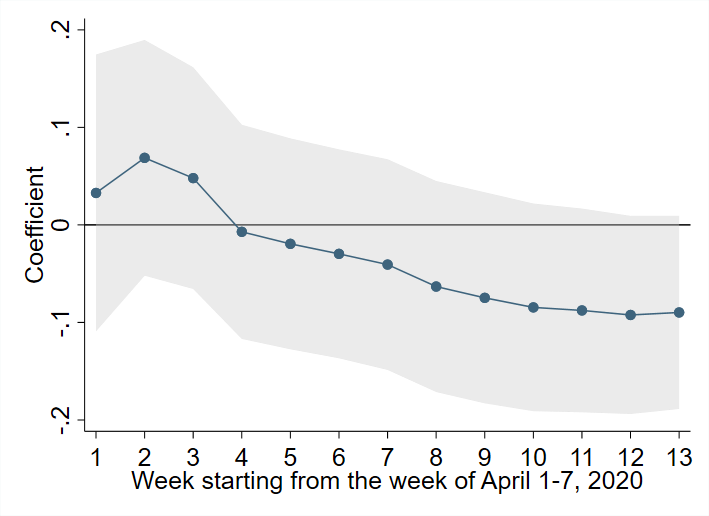}
		\label{fig: total_no_hins_deaths}
		\subcaption{No health insurance}
	\end{subfigure}%
	\begin{subfigure}{.166\textwidth}
		\centering
		\includegraphics[width=\linewidth]{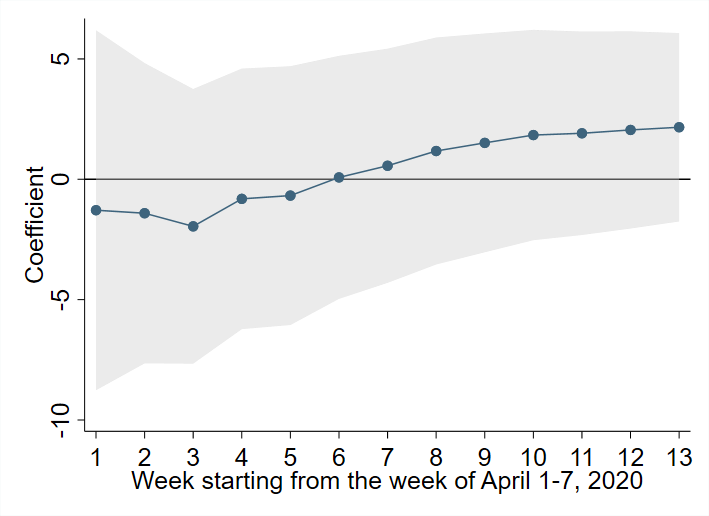}
		\label{fig: log_med_income_deaths}
		\subcaption{Log median income}
	\end{subfigure}%
	\begin{subfigure}{.166\textwidth}
		\centering
		\includegraphics[width=\linewidth]{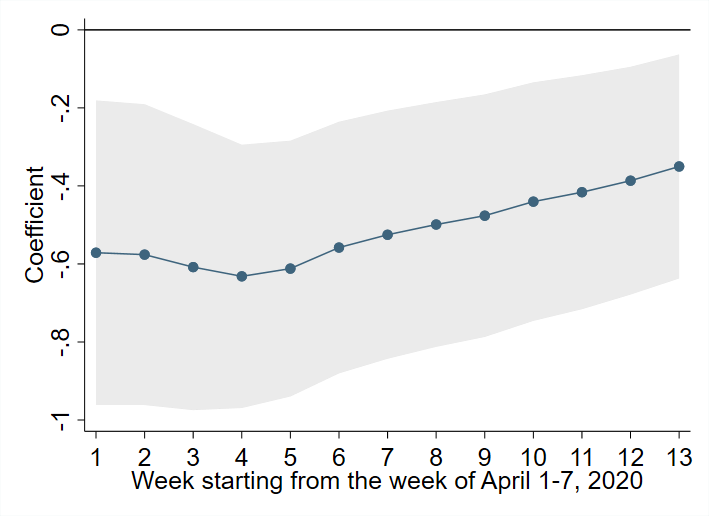}
		\label{fig: age_20_39_deaths}
		\subcaption{Share of age 20-39}
	\end{subfigure}
	\begin{subfigure}{.166\textwidth}
		\centering
		\includegraphics[width=\linewidth]{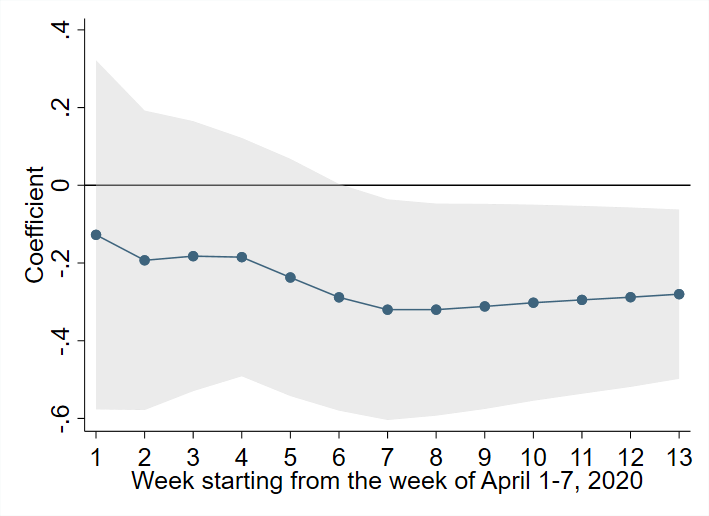}
		\label{fig: age_40_59_deaths}
		\subcaption{Share of age 40-59}
	\end{subfigure}%
	\begin{subfigure}{.166\textwidth}
		\centering
		\includegraphics[width=\linewidth]{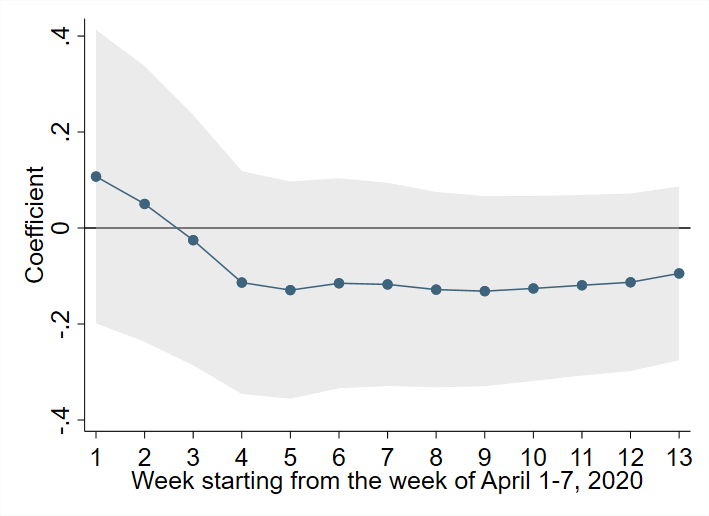}
		\label{fig: age_60_plus_deaths}
		\subcaption{Share of age 60+}
	\end{subfigure}%
	\begin{subfigure}{.166\textwidth}
		\centering
		\includegraphics[width=\linewidth]{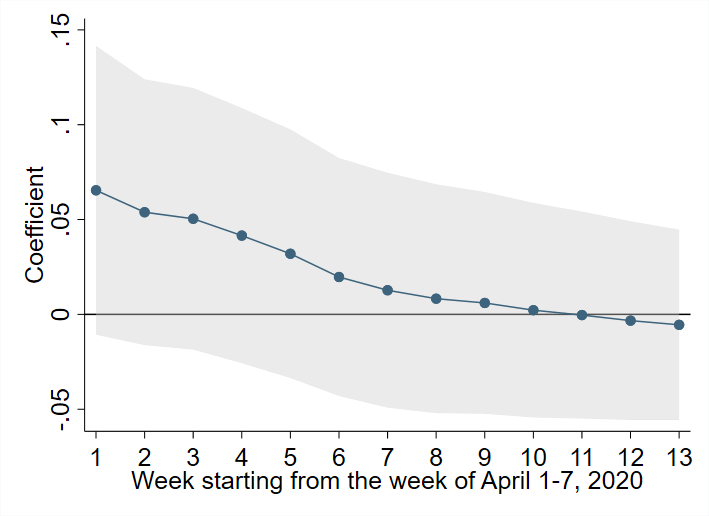}
		\label{fig: black_share_deaths}
		\subcaption{Share of Black}
	\end{subfigure}%
	\begin{subfigure}{.166\textwidth}
		\centering
		\includegraphics[width=\linewidth]{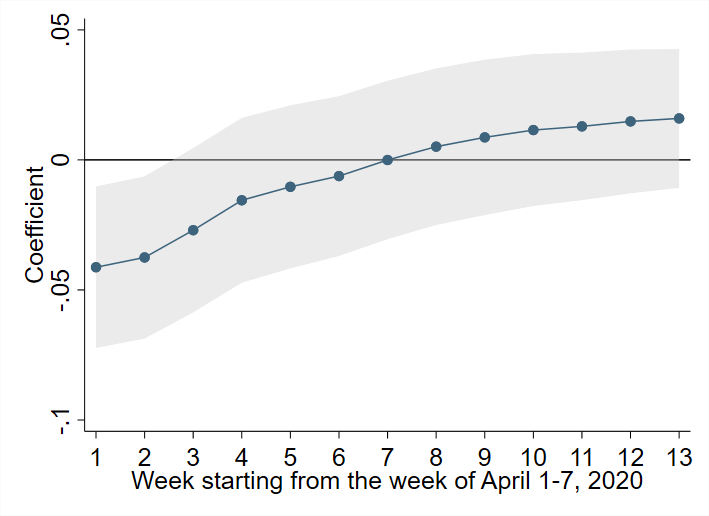}
		\label{fig: hispanic_share_deaths}
		\subcaption{Share of Hispanic}
	\end{subfigure}%
	\begin{subfigure}{.166\textwidth}
		\centering
		\includegraphics[width=\linewidth]{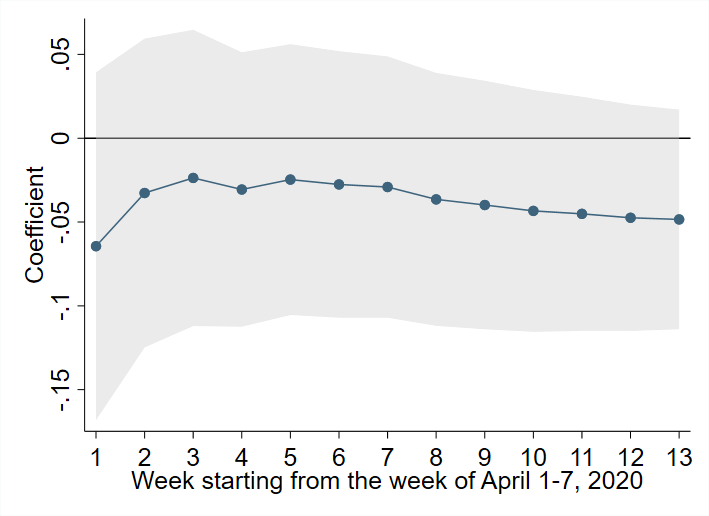}
		\label{fig: asian_share_deaths}
		\subcaption{Share of Asian}
	\end{subfigure}%
	\begin{subfigure}{.166\textwidth}
		\centering
		\includegraphics[width=\linewidth]{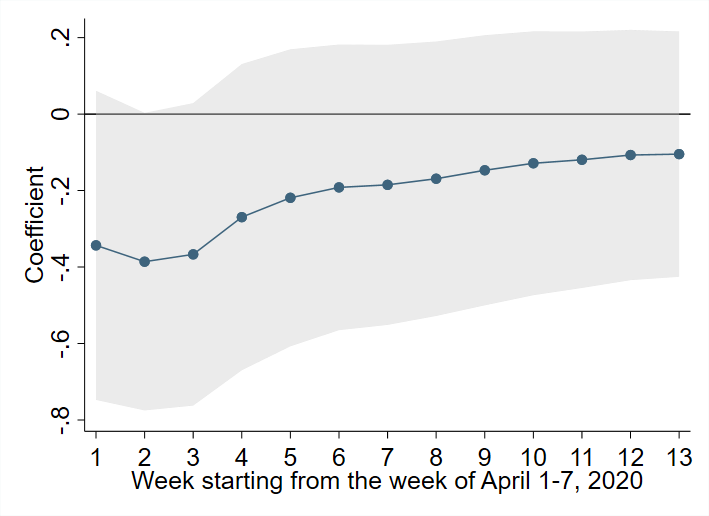}
		\label{fig: occ_1_deaths}
		\subcaption{Ess-Professional}
	\end{subfigure}
	\begin{subfigure}{.166\textwidth}
		\centering
		\includegraphics[width=\linewidth]{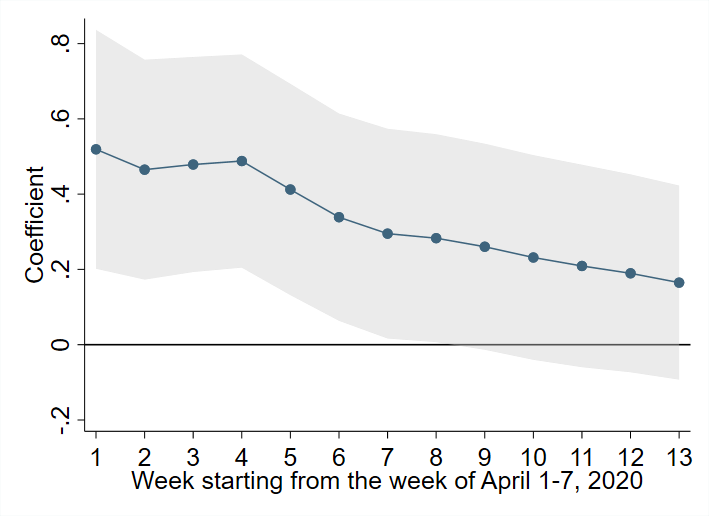}
		\label{fig: occ_2_deaths}
		\subcaption{Non-Ess-Prof}
	\end{subfigure}%
	\begin{subfigure}{.166\textwidth}
		\centering
		\includegraphics[width=\linewidth]{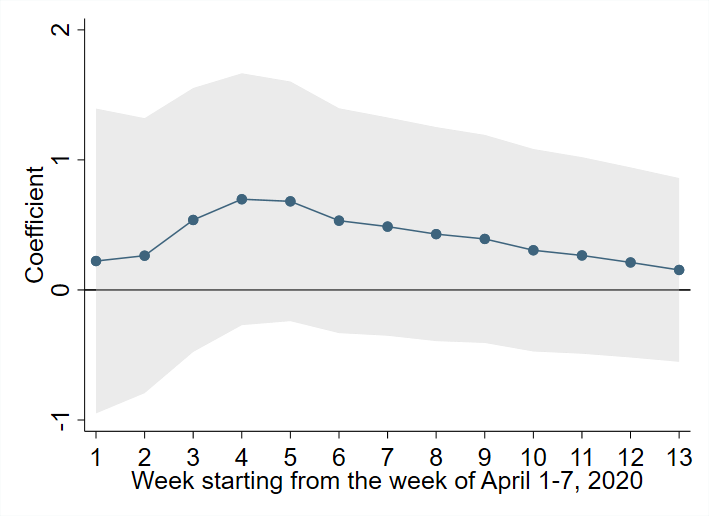}
		\label{fig: occ_3_deaths}
		\subcaption{Science}
	\end{subfigure}%
	\begin{subfigure}{.166\textwidth}
		\centering
		\includegraphics[width=\linewidth]{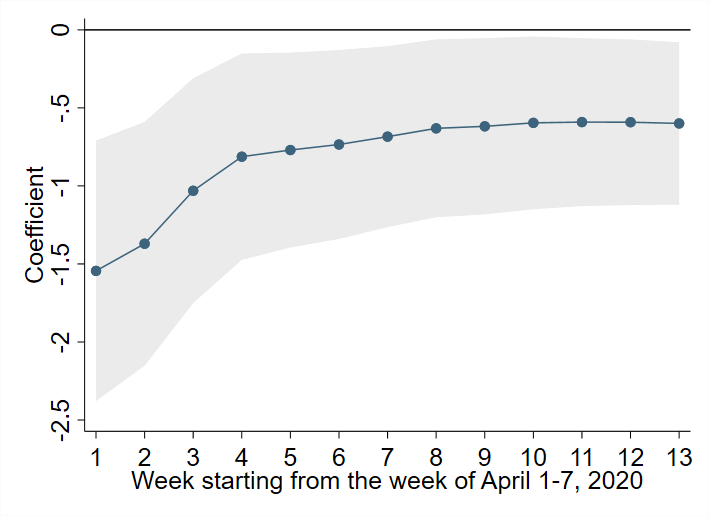}
		\label{fig: occ_4_deaths}
		\subcaption{Law and Related}
	\end{subfigure}%
	\begin{subfigure}{.166\textwidth}
		\centering
		\includegraphics[width=\linewidth]{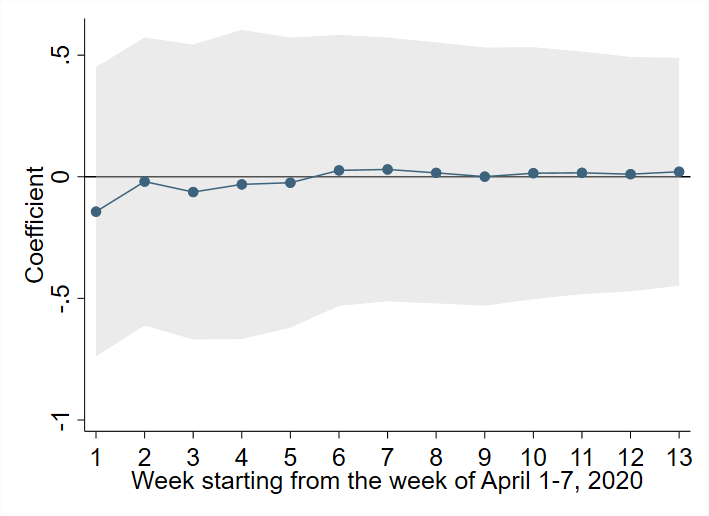}
		\label{fig: occ_5_deaths}
		\subcaption{Health Practitioners}
	\end{subfigure}%
	\begin{subfigure}{.166\textwidth}
		\centering
		\includegraphics[width=\linewidth]{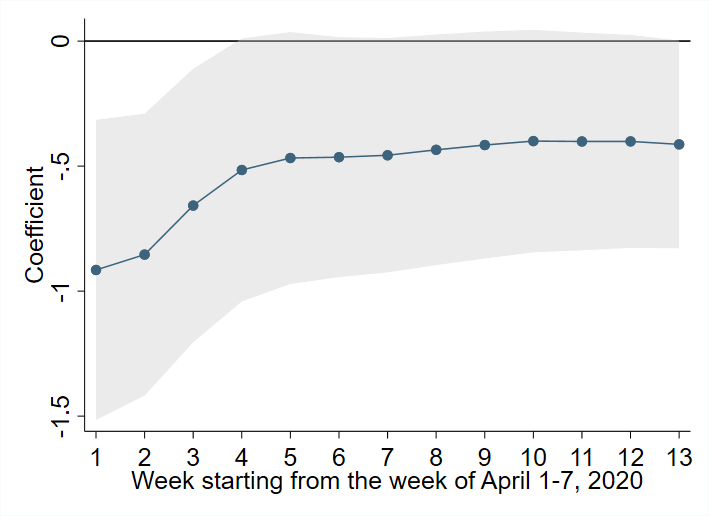}
		\label{fig: occ_6_deaths}
		\subcaption{Other Health}
	\end{subfigure}%
	\begin{subfigure}{.166\textwidth}
		\centering
		\includegraphics[width=\linewidth]{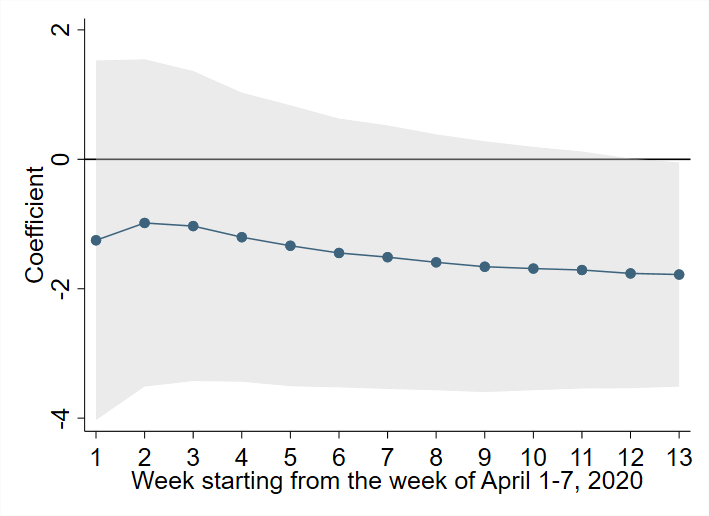}
		\label{fig: occ_7_deaths}
		\subcaption{Firefighting}
	\end{subfigure}
	\begin{subfigure}{.166\textwidth}
		\centering
		\includegraphics[width=\linewidth]{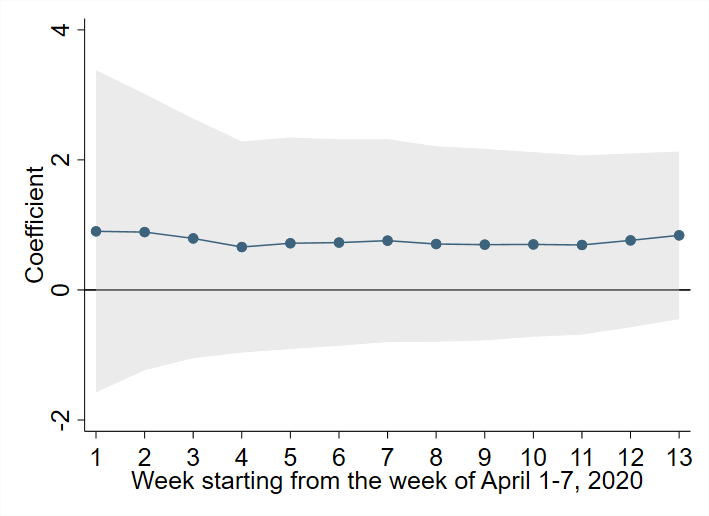}
		\label{fig: occ_8_deaths}
		\subcaption{Law Enforcement}
	\end{subfigure}%
	\begin{subfigure}{.166\textwidth}
		\centering
		\includegraphics[width=\linewidth]{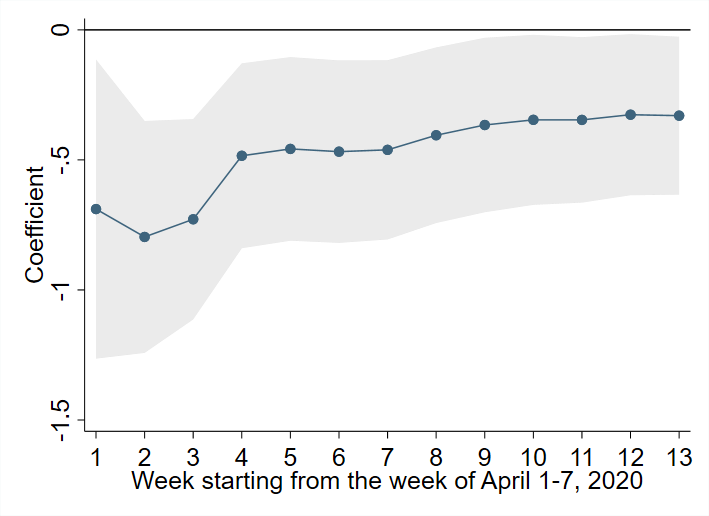}
		\label{fig: occ_9_deaths}
		\subcaption{Essential-Service}
	\end{subfigure}%
	\begin{subfigure}{.166\textwidth}
		\centering
		\includegraphics[width=\linewidth]{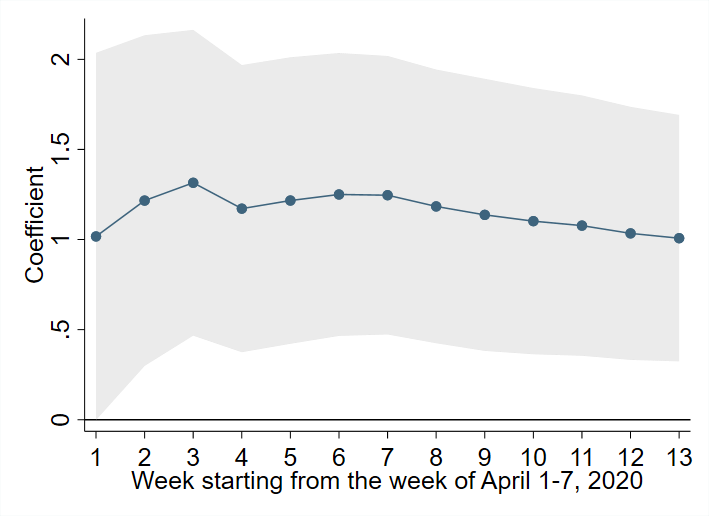}
		\label{fig: occ_10_deaths}
		\subcaption{Non-Ess-Service}
	\end{subfigure}%
	\begin{subfigure}{.166\textwidth}
		\centering
		\includegraphics[width=\linewidth]{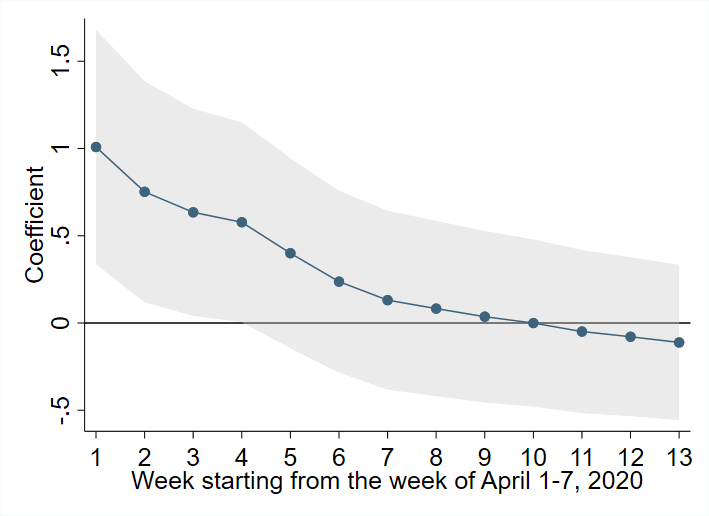}
		\label{fig: occ_11_deaths}
		\subcaption{Industrial \& Constr.}
	\end{subfigure}%
	\begin{subfigure}{.166\textwidth}
		\centering
		\includegraphics[width=\linewidth]{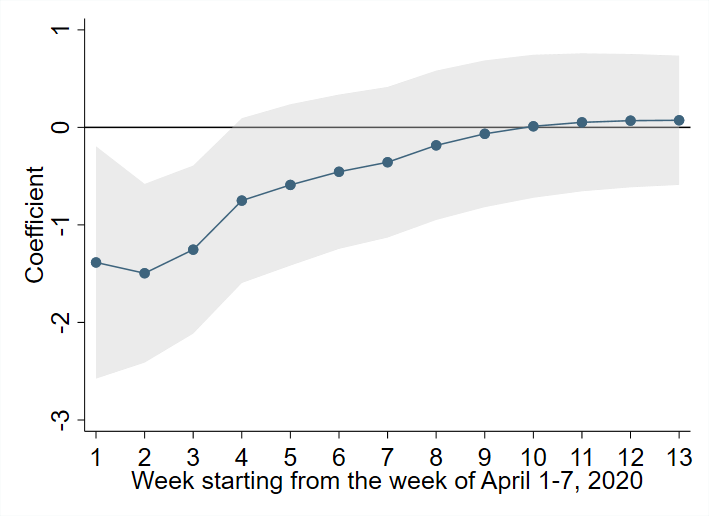}
		\label{fig: occ_12_deaths}
		\subcaption{Essential-Technical}
	\end{subfigure}%
	\begin{subfigure}{.166\textwidth}
		\centering
		\includegraphics[width=\linewidth]{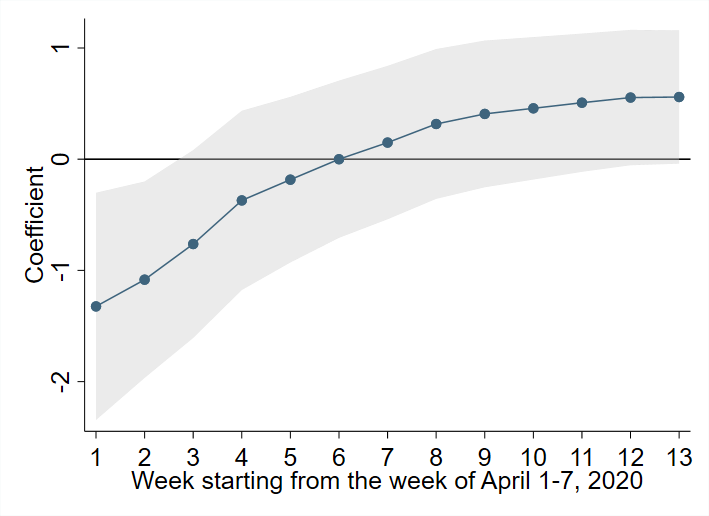}
		\label{fig: occ_13_deaths}
		\subcaption{Transportation}
	\end{subfigure}%
	\caption{Weekly evolution of coefficients (dependent variable: log of daily cumulative deaths per 100000 inhabitants)}
	\label{fig: coeff_deaths_other}
	\begin{spacing}{1.0}\justify\footnotesize{\textsc{Notes}: The figure shows the impact of state characteristics on COVID-19 deaths. Coefficients are from OLS regression \eqref{eq: regression_deaths} for the weeks from April 1-7, 2020 to June 24-July 1, 2020. The shaded area represents the 95\% confidence interval.}\end{spacing}
\end{figure}
\end{sidewaysfigure}

\end{document}